# Global Universal Scaling and Ultra-Small Parameterization in Machine Learning Interatomic Potentials with Super-Linearity

**Authors:** Yanxiao Hu [1†], Ye Sheng [1†], Jing Huang [1], Xiaoxin Xu [1,2], Yuyan Yang [2], Mingqiang Zhang [1], Yabei Wu [1,3], Caichao Ye [1,3], Jiong Yang [2] and Wenqing Zhang [1,3*]

**Affiliations:**

[1] Department of Materials Science and Engineering, Southern University of Science and Technology, Shenzhen, Guangdong 518055, China

[2] Materials Genome Institute, Shanghai Engineering Research Center for Integrated Circuits and Advanced Display Materials, Shanghai University, Shanghai 200444, China.

[3] Institute of Innovative Materials & Guangdong Provincial Key Lab for Computational Science and Materials Design, Southern University of Science and Technology, Shenzhen, Guangdong 518055, China

*Corresponding author. Email: zhangwq@sustech.edu.cn

†These authors contributed equally to this work.

# Abstract

Using machine learning (ML) to construct interatomic interactions and thus potential energy surface (PES) has become a common strategy for materials design and simulations. However, those current models of machine learning interatomic potential (MLIP) provide no relevant physical constrains, and thus may owe intrinsic out-of-domain difficulty which underlies the challenges of model generalizability and physical scalability. Here, by incorporating physics-informed *Universal-Scaling* law and nonlinearity-embedded interaction function, we develop a *Super-linear* MLIP with both *Ultra-Small* parameterization and greatly expanded expressive capability, named SUS$^2$-MLIP. Due to the global scaling rooting in universal equation of state (UEOS), SUS$^2$-MLIP not only has significantly-reduced parameters by decoupling the element space from coordinate space, but also naturally outcomes the out-of-domain difficulty and endows the potentials with inherent generalizability and scalability even with relatively small training dataset. The nonlinearity-enbeding transformation for interaction function expands the expressive capability and make the potentials super-linear. The SUS$^2$-MLIP outperforms the state-of-the-art MLIP models with its exceptional computational efficiency especially for multiple-element materials and physical scalability in property prediction. This work not only presents a highly-efficient universal MLIP model but also sheds light on incorporating physical constraints into artificial-intelligence-aided materials simulation.

# I. Introduction

The potential energy surface (PES) and interatomic potentials play a critical role in large-scale atomistic simulations. It depicts energy for thermodynamics, forces for dynamics and kinetics, and higher-order anharmonic interactions for lattice vibrations (phonons) and thermal transports etc. The traditional approaches to PES construction primarily consist of accurate yet scale-limited first-principles electronic structure calculations, and efficient but precision-limited empirical force fields fitting to known fundamental physical principles. However, the recent emergence of machine learning interatomic potentials[1–5] (MLIPs) presents a paradigm-shifting solution to this long-standing trade-off dilemma.

MLIPs aim to accurately fit the high-dimensional PES of complex materials by employing machine learning (ML) techniques in conjunction with small-scale density functional theory (DFT) calculations. With continuous refinement and improvement over the past decades, MLIPs have demonstrated remarkable precision and efficiency for simulations across various material systems at multiple scales. Empowered by leading neural network framework, the reliability of state-of-the-art MLIP models (e.g., MACE[6], Allegro[7] and CHGNet[8] el.) could even approach that of DFT calculations in energy-related simulations, but the MLIP exhibit much greater computational efficiency than the DFT-based methods. MLIPs present an alluring alternative to the burdensome DFT calculations in the simulations and computational design for catalyst[9,10], battery[11–13], new compounds[14–16], and so on.

Despite the remarkable achievements in various research fields, MLIPs still encounter significant challenges. These challenges can be summarized into two main areas: generalizability[3] and physical scalability[17,18]. The first stems from the inherent conflict between prevailing philosophical perspectives on ML and the core principles of physical modeling. As a data-driven approach, ML aims to free complex modeling from human cognitive limitations and traditional interpretability norms by leveraging statistical analysis of big data for predictions about the real world[19]. However, due to the intrinsic interpolation nature of ML methods, models often struggle with extrapolation when faced with out-of-distribution data. This limitation poses a substantial challenge to the reliability of MLIPs in complex dynamic simulations, raising the second issue regarding how to ensure the physical scalability of MLIPs that are trained on energy and atomic forces for predicting macroscopic properties. To address these dual challenges, a synergistic integration of physical insights with ML architectures becomes imperative. Specifically, the development of MLIP necessitates 1) a comprehensive understanding of the fundamental governing equations underlying interatomic interactions, combined with 2) the incorporation of appropriate global scaling to ensure both model parsimony[20] and transferability across diverse material systems.

In this study, we propose a MLIP model in generalized linear framework, termed SUS$^2$-MLIP, which is rigorously derived from the principles of many-body physics and presented with an analytical expression. Drawing inspiration from the physical constraints underlying the well-established universal equation of state[21] (UEOS), we introduce the global *universal scaling* for radial function. The universal radial function describes all types of atom-atom interaction in principle, addressing the exponential increase in parameters and data requirements resulting from the increase in elemental diversity. The accuracy of SUS$^2$-MLIP is assessed across various databases including semiconductors, alloys, perovskite oxides, and other materials. While only utilizing (2-3)-orders-of-magnitude less parameters, named *ultra-small* parameterization, our framework achieves comparable performance to other state-of-the-art models that rely on complex deep neural network or graph neural network architectures. Within such a compact parameter space, SUS$^2$-MLIP still enables the construction of physically scalable PES, thereby facilitating reliable simulations for macroscopic material properties including ion diffusion and thermal transport with high-order anharmonic interactions.

## II. Theoretical Framework

**II.1 General formulation of potential energy surface as a property field**

Let us define PES or a physical property as an effective field at first, and then derive our MLIP model in a strictly analytical way. In principle, by dividing a group of atoms into clusters, the global property could be obtained by summing over that of all clusters by assuming a limited range of physical interactions. For the cluster with central atom-*I* surrounding by neighbors ($j$ = 1, …, N), named as cluster-*I*, the property of the cluster-*I* ($P_I$) can be reasonably approximated as the following expression:

$$P_I(\mathbf{r_I}; \{\mathbf{r_j}\}) = p_I \prod_{j \neq I} (\varphi_{Ij}(\mathbf{r_{Ij}}, Z_I, Z_j) + 1) \qquad (1)$$

where the coordinate vector $\mathbf{r_{Ij}}$ of the $j^{th}$ atom in cluster-*I* is defined as $\mathbf{r_{Ij}} \equiv \mathbf{r_j} - \mathbf{r_I}$ referring to the central atom *I*. $Z_I(Z_j)$ denote elemental type of atom *I* or *j*. The $\varphi_{Ij}(\mathbf{r_{Ij}}, Z_I, Z_j)$ could be considered as a function describing the contribution of atom-*j* to the property of cluster-*I*, and equivalently it defines the effective interaction between the central atom *I* and nearby atom *j*. Note that $\varphi_{Ij}(\mathbf{r_{Ij}}, Z_I, Z_j) \to 0$ as $\mathbf{r_{Ij}}$ exceeds the interaction range. Therefore, $p_I$ could be considered as an integrating quantity describing the property of atom-*I*-centered cluster. Physically, equation (1) describes the property of a cluster with many interacting atoms, in analog to the wave functions in quantum many-body system. In this aspect, equation (1), together with its summation of clusters, define an effective physical property field under the mean-field approximation (refer to *Supplementary Information S.1*), as illustrated in Fig. 1. Note that all neighboring atoms engage in interacting solely with the central atom-*I* within cluster-*I*, while the complex indirect interactions could be automatically included by treating every atom with surroundings as a unique cluster in parallel and thus the summation over *I*.

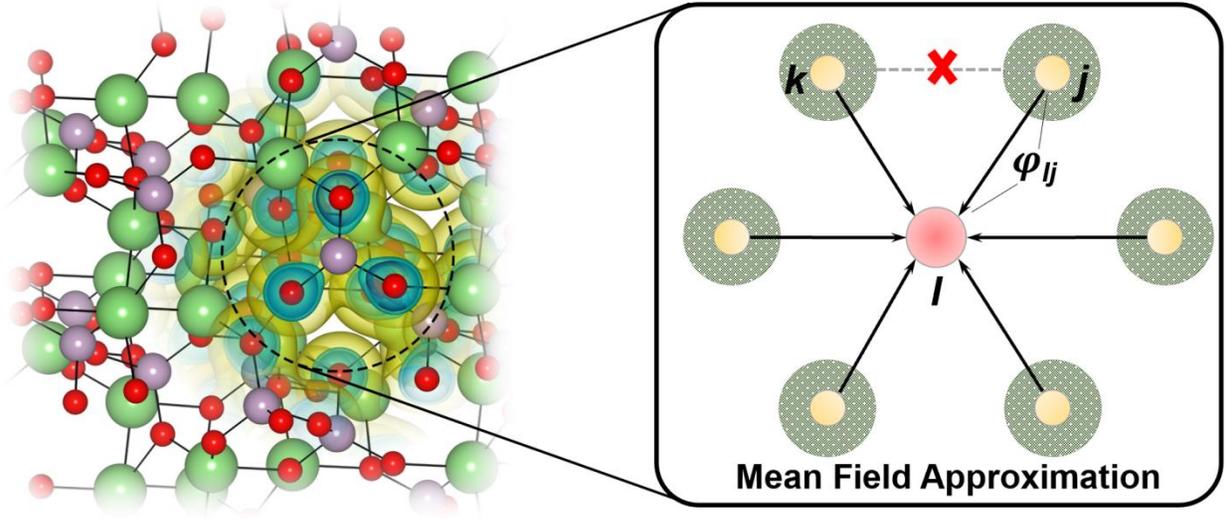

**Fig. 1 Property field under mean field approximation.** Physically, the property field $P_I$ describes the local property distribution of the atom-$I$-centered cluster with many atoms in analog to the wave functions in quantum many-body system. $P_I$ and its summation over $I$ define a global effective field depending on the relative distances between atoms as well as on the expansion/compression of the whole system.

The property field $P_I(r_I; \{r_j\})$ in equation (1) varies continuously as the interatomic distances change or the whole system undergoes volume expansion. This naturally leads to the implementation of physics-informed universal scaling law into function $\varphi_{Ij}$, and the scaling originates in the well-established UEOS of materials[21]. Details will be discussed in *II.2*. Meanwhile, equation (1) is also consistent with the conventional decomposition of potential energy surface by expanding it as:

$$P_I = p_I \left( 1 + \sum_j \varphi_{Ij} + \sum_{j<j'} \varphi_{Ij}\varphi_{Ij'} + \sum_{j<j'<j''} \varphi_{Ij}\varphi_{Ij'}\varphi_{Ij''} + \cdots \right) \quad (2)$$
$$= p_I \left( 1 + \varphi_I + \frac{1}{2}\varphi_I^2 + \frac{1}{3!}\varphi_I^3 + \cdots \right)$$

Such an expression is highly like that used in other MLIPs such as atomic cluster expansion[22,23] (ACE), moment tensor potential[24] (MTP), etc. In equation (2), $\varphi_I = \sum_j \varphi_{Ij}$ is the collection of pair contribution of all neighbors, the term $\varphi_I^2 = \sum_{j<j'} \varphi_{Ij}\varphi_{Ij'}$ contains all *j-I-j'* 3-body interactions, and $\varphi_I^n$ describes all the (n+1)-body interaction in the cluster. The difference of equation (2) from the conventional decomposition is the appearance of an atom-*I*-related effective quantity $p_I$ and the concise expression for all many-body interactions that intrinsically share the same function $\varphi_I = \sum_j \varphi_{Ij}$. As will be seen, the property field approximation makes the current model very concise and also MLIP development analytically tractable.

As the specific property refers to the field of potential energy, e.g. the $E_I$, equation (2) should be expressed in the following form:

$$E_I = \varepsilon_{Z_I}\left(1 + <\varphi_I>_0 + \frac{1}{2}<\varphi_I^2>_0 + \frac{1}{3!}<\varphi_I^3>_0 + \cdots\right) + \widetilde{\varepsilon_{Z_I}} \tag{3}$$

The $<\varphi_I^n>_0$, with the subscript zero, signifies extracting the whole scalar space of the (n+1)-body interaction, which respects the fact that energy is invariant under the operations of three-dimensional Euclidean group *E(3)*. Note that the permutation invariance is automatically kept in equation (1) and thus equations (2-3). $\varepsilon_{Z_I}$ is the cluster energy scale and $\widetilde{\varepsilon_{Z_I}}$ the reference energy, respectively. In principle, the construction of PES is deemed complete once $\varepsilon_{Z_I}$, $\widetilde{\varepsilon_{Z_I}}$ and $\varphi_{Ij}$ are determined.

## II.2 Physics-informed Universal Scaling for global scalability

The function $\varphi_{Ij}(\mathbf{r}_{Ij}, Z_I, Z_j)$ defines the effective interaction between the atom *I* and nearby atom *j*, and is expanded on a complete basis set $\{\mathcal{B}\}$, which read $\varphi_{Ij}(\mathbf{r}_{Ij}, Z_I, Z_j) = \sum_{\tilde{i}} C_{\tilde{i}} \mathcal{B}_{\tilde{i}}(\mathbf{r}_{Ij}, Z_I, Z_j)$. $\mathcal{B}_{\tilde{i}}$ consists of a product of a radial function and angular basis function. Either the spherical harmonics $\{Y_l^m\}$ or $\{\hat{r}^{\otimes l}\}$ in the directional unit vector $\hat{r}_{Ij}$ is commonly chosen as a set of angular basis function.

In our model, we firstly re-express the functional basis set $\{\mathcal{B}\}$, by introducing two addition parameters $t_I$ and $t_j$, as,

$$\mathcal{B}_{\tilde{i}}(\mathbf{r}_{Ij}, Z_I, Z_j) = t_I * R_l(r_{Ij}, Z_I, Z_j) * A_{\tilde{i}}(\hat{r}_{Ij}) * t_j \tag{4}$$

Here $t_I$ ($t_j$) relates to atom $I$ or atom $j$ and could be reasonably considered as atom-type or element-specific quantity. In other word, equation (4) tends to normalize the $I$-$j$ pair interaction with the element information from the specific atoms $I$ and $j$, so that the radial function, $R_l$, could be much general and solely depends on elemental type. $A_{\vec{l}}$ ($\{Y_l^m\}$ or $\{\hat{r}^{\otimes l}\}$) is angular function as discussed. The subscript $l$ denotes a representation space of the three-dimensional rotation group $SO(3)$, and $\vec{l}$ refers to a specific component of this subspace, e.g., $(lm)$ for spherical harmonics. As usual, the radial function $R_l$ could be expanded on a complete set of radial basis functions $\{T_n\}$, e.g. Bessel functions or Chebyshev polynomials.

Up to now, the approach in equations (1-4) does not take into account of any physical constraint, similar to all other state-of-the-art MLIP models. The lack of global scaling could be an important reason that leads to vast parameter space for acceptable representation currently. A typical example is the exponential parameter growth arising from the increase in the elemental diversity ($R_l = \sum_n C_n^{l,Z_I Z_j} T_n$ with $C_n^{l,Z_I Z_j} \in \mathbb{R}^{n \times l \times Z^2}$) in developing universal potentials[6,8,25–30] for multiple-element systems. Here, we prove that, by using the property field concept and equation (1), there exists an intrinsic and universal scaling for radial function $R_l(r)$ for all-type pair interaction, originating from the well-established knowledge of UEOS. This global scaling leads to substantial reduction of the parameters.

The UEOS, recognized for a long time, states that the dependence of normalized cohesive energy, i.e. the potential energy of an interacting system, on the changing volume or lattice constant follows a universal equation after scaling interatomic distance (or lattice constant) by only two system-dependent or element-specific effective parameters. In simple, there exists a simple yet universal equation as,

$$E^* \equiv \frac{E}{E_0} = -(r^* + 1)e^{-r^*}, r^* = \alpha(\frac{r}{r_0} - 1) \qquad (5)$$

where $E^*$ is the normalized cohesive energy, $r^*$ the scaled length or distance. $\alpha$ and $r_0$ are the system-dependent scaling factor. Because of the dimensionless nature of equation (5), it can be applicable across various scales. It indicates that all types of interatomic interaction are mapped onto a shared space after global scaling, which significantly simplifies the mathematical formulation of the model. The universality of equation (5) has been proven to be valid for all atomic systems[31], encompassing crystallines, molecules, atom pairs, adhered interfaces, and even nuclear-nuclear interaction.

The property field in equation (1) and the scaling law in equation (5) lead to natural implementation of the EOS-based universal scaling or physics-informed constrain into the present model. By simply applying equation (1) to one-element FCC or BCC crystal, the property field $P_I(\boldsymbol{r_I};\{\boldsymbol{r_j}\})$ reduces to a simple one-dimensional function of the form $P_I(\boldsymbol{r_{nn}})\sim r_{nn}$, in which $r_{nn}$ is the nearest-neighbor interatomic distance. The $P_I(\boldsymbol{r_{nn}})$ is actually the equation of state of the isotropic system and has to satisfy the UEOS. Therefore, all $\varphi_{Ij}$ should also possess the universal representation within the scaling physical space as in equation (5). Considering an atom pair, as shown in Fig. 2(a), the radial part of the interaction function of the atom-pair could be firstly normalized to be elemental-type-dependent, and secondly the element-type dependent $R_l$ could be further expressed in the element-type-irrelevant universal radial function $\tilde{R}_l(r_{Ij}^*)$ after the universal scaling, i.e.,

$$R_l \to \tilde{R}_l = \tilde{R}_l(r_{Ij}^*), r_{Ij}^* = \alpha_{Z_IZ_j}(\frac{r_{Ij}}{r_0^{Z_IZ_j}} - 1) \qquad (6)$$

Thus, $\tilde{R}_l(r_{Ij}^*)$ serves as a general and universal description of interactions for all pairs. This is important for allowing the radial expansion coefficients to decouple the coordinate space from the elemental space. It has an important consequence that such a universal expression, equation (6), could reduce much the parameter space from $\mathbb{R}^{n\times l\times Z^2}$ to $\mathbb{R}^{n\times l}$ ($\tilde{R}_l = \sum_n \tilde{C}_n^l T_n$). Essentially, a MLIP has to satisfy equation (5) when systems experience uniform volumetric expansion or compression, ensuring the capability of the potentials to effectively describe systems that deviate significantly from equilibrium. In other word, this stringent physical constraint enables the current MLIPs, even trained on a limited amount of training data in proximity to equilibrium, to effectively generalize into the out-of-equilibrium regions as illustrated in Fig. 2(b). Due to the element-type-independent nature of $\tilde{R}_l$, the current SUS$^2$-MLIPs can also be trained using diverse types of data and still maintain its effectiveness, thereby significantly enhancing the model's capacity for data utilization even into the far-from-equilibrium states with strong anharmonicity.

Notably, equation (4) exhibits a node-edge-node structure similar to that of graph neural networks, where $t_I$ and $t_j$ corresponds to the information of nodes I and j, and pair function to node interactions. The distinction lies in the fact that, node information in graph neural networks is expressed as a long vector that could be updated with graph convolution. Under the mean field approximation, $t_I$ is an effective parameter related to the element type of atom I. It is worth addressing that while $t_I$ and $t_j$ looks like fitting parameters in equation (4), they could also be used for real element-property embedding for advanced MLIP with a straightforward extension of the above equations (see *Supplementary Information S.2*).

Building upon the aforementioned theoretical framework, we proposed SUS$^2$-MLIP model which is realized through the modified *mlip*[32] package. The technical details regarding to the selection of basis functions, the sets of scaling parameters, the construction of universal radial functions, and model regression methodologies can be found in *Supplementary Information S.2*.

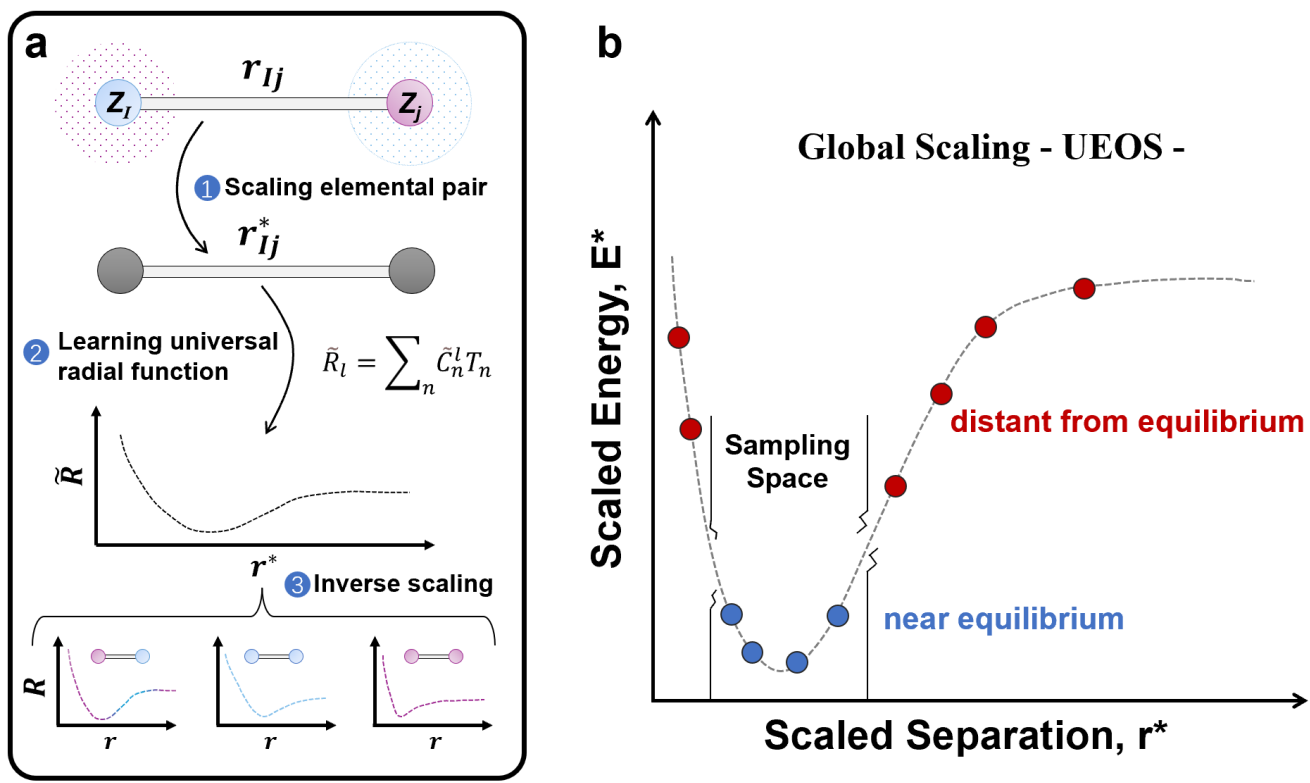

**Fig. 2 Construction of universal radial function with the Physics-informed Universal scaling. (a)** The elemental pairs are mapped, after scaling, to an element-independent $r^*$ space to learn the universal radial function; ultimately real-space radial function of atomic pairs can be obtained by inverse mapping. **(b)** The inherent physical constraint from UEOS endows the current potential model with enhanced generalizability and scalability, even with only the few data around equilibrium (blue points within the walls), for far-from-equilibrium points (in red).

### II.3 Nonlinearity-embedded radial function for Super-Linear expressive capacity

Those state-of-the-art MLIP models often adopt the basis of linear combination of radial functions, usually in a simplified form. The GNN-based MLIPs rely on the complex neural-network-layer-convolution to enhance the model capability, leading to the problems of huge parameter space and interpretation. To expand the expressive capacity of the present SUS$^2$-MLIP, we follow the idea of adaptive memory[33] of Hopfield network to incorporate nonlinear transformation for coordinate function into basis, which endows models with super-linear memory storage capacity.

The traditional Hopfield networks[34] exhibit linear scaling relationships for storing memory patterns that hardly exceed the dimensionality of the feature space. In a many-body system, the existence of many hidden interaction patterns beyond the mean-field approximation, together with the nonlinear interactions growing rapidly with body-orders, thereby amplifies the significance of reliable radial functions and basis. To overcome the linear scaling memory challenge, a strategy, that is the adaptive memory approach, is to introduce activation function $\sigma(\cdot)$ to enhance expressive capability. In a similar or equivalent way, to expand the capability of the current model, the radial function for atom-pair interaction is expressed in the following composite mapping:

$$R(r) = \tilde{R}(x) = \tilde{R}(\sigma(r)) \qquad (8)$$

where $\tilde{R}$ is the generalized radial function with the argument $x$ obtained by the continuous nonlinear mapping $\sigma = \sigma_1 \circ \sigma_2 ... \circ \sigma_n$. Equation (8) corresponds to the transformation from traditional Hopfield network ($E = -\sum_\mu (\xi_\mu . x)^2$) to the dense associative memory model ($E = -\sum_\mu \sigma(\xi_\mu . x)$). When $\sigma$ grows more rapidly more than quadratic, model can achieve a super-linear memory storage capacity. It is also worth noting that the introduction of nonlinear patterns in this context aligns formally with the Kolmogorov-Arnold-Network[35] (KAN) models. Essentially, when the form of $\sigma$ is appropriately selected, the model scaling law will be conspicuously enhanced. KAN achieves this by adjusting the learnable activation function.

The introduction of nonlinear transformations, equation (8), implies the model capability expansion of embedding nonlinearity into the basis function, thus named nonlinearity-embedded radial function approach. As proved in adaptive memory approach and KAN network, the capability enhancement is substantial and even close to be exponential, making the current model very much super-linear in theory. Note that the current approach differs from neural network models based on traditional multilayer perceptron (MLP). In MLP, the model's expressive capability necessitates the superposition of multiple sets of nonlinear functions, which results in the generation of a substantial number of parameters. Interestingly, the incorporation of the nonlinearity-embedding in our framework is an extremely natural process. For instance, the mapping from $r$ to $r^*$ is an intrinsic physical constraint (equation (6)), and the nonlinear mapping from $r^*$ to $x \in (-1,1)$, *tanh* (equation S2 in *Supplementary Information S.2*), is a rational choice that satisfies both the Chebyshev polynomial and the fundamental image of atomic interaction. As an exponential derived function, *tanh* enables $\tilde{R}$ to possess a super-polynomial[36] expansion form, thereby endowing SUS$^2$-MLIP with super-linear expressive capacity.

## III. Applications and Results

### III.1 Ultra-small parameter space and general benchmarks

In this section, SUS$^2$-MLIP is compared to the popular state-of-the-art neural network MLIP models, including GemNet-OC[37] (GNO), EquiformerV2[38] (EFV2), Nequip[39], Allegro[7], and DPA-2[40]. To evaluate the accuracy of SUS$^2$-MLIP in learning atomic energy and force across various systems, this benchmark includes datasets covering alloys, cathode materials, perovskite oxides, and semiconductors. More details about the datasets and corresponding model training can be found in the *Methods* section.

The primary findings, as shown in Fig. 3, demonstrate that the SUS$^2$-MLIP achieves comparable accuracy to other leading MLIP models. These datasets share a common characteristic of exhibiting element diversity. For instance, the *alloy* dataset consists of 53 elements, each capable of coexisting with others through metallic bonding. This highlights the effectiveness of the "universal radial function" in describing interactions within complex elemental space. Based on results in *Cathode*, *Perovskite*, and *semiconductor* modeling, SUS$^2$-MLIP exhibits evident scalability across polar and covalent systems. Notably, SUS$^2$-MLIP accomplishes these outcomes with significantly fewer parameters compared to other models. For example, DPA-2, the latest DeepMD[41] architecture, has over 5 million parameters, while the parameter count of SUS$^2$-MLIP for *alloy, cathode*, *perovskite* and *semiconductor* modeling is merely 17,689, 2,130, 6,092 and 6,786 respectively.

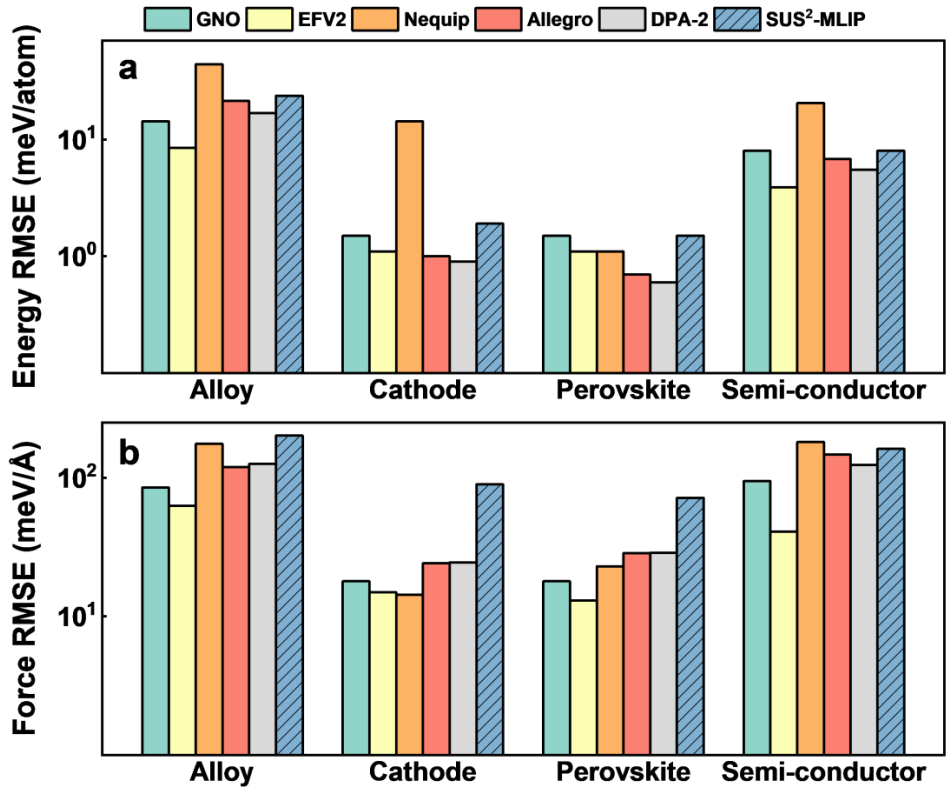

**Fig. 3 General benchmark. (a)** Energy and **(b)** force root-mean-square error (RMSE) on the datasets of alloy, cathode, perovskite and semiconductor. Results of other neural network MLIP models are reported in ref[40]. The training and validation datasets for *semiconductors* in this work were chosen based on a criterion of force < 5 eV/Å.

To evaluate the performance of SUS$^2$-MLIP in an extremely complex sample space, we trained it on the complete periodic table using the same dataset (excepting for the noble gases) of M3GNet[25]. The SUS$^2$-MLIP achieved a mean absolute error (MAE) of 63 meV/atom for energy, 155 meV/Å for force, and 1.0 GPa for stress. Although M3GNet outperformed SUS$^2$-MLIP with lower MAEs (35 meV/atom, 72 meV/Å, and 0.41 GPa), its parameter count (227,549) is an order of magnitude higher compared to that in SUS$^2$-MLIP (29,006), suggesting that our approach demonstrates superior parameter efficiency. To further highlight this advantage, we present a comparative analysis of parameter counts between SUS$^2$-MLIP with MTP (a popularly used MLIP model with a linear framework) in Fig. 4. As the diversity of element types increases, the lightweight nature of SUS$^2$-MLIP becomes more evident. The inherent parsimony ensures its superior computational efficiency and enables potential applications in larger-scale simulations.

The training results of additional datasets, including half-Heusler materials, Cu$_2$Se, and sulfide solid-state electrolytes, can be found in *Supplementary Information S.3*.

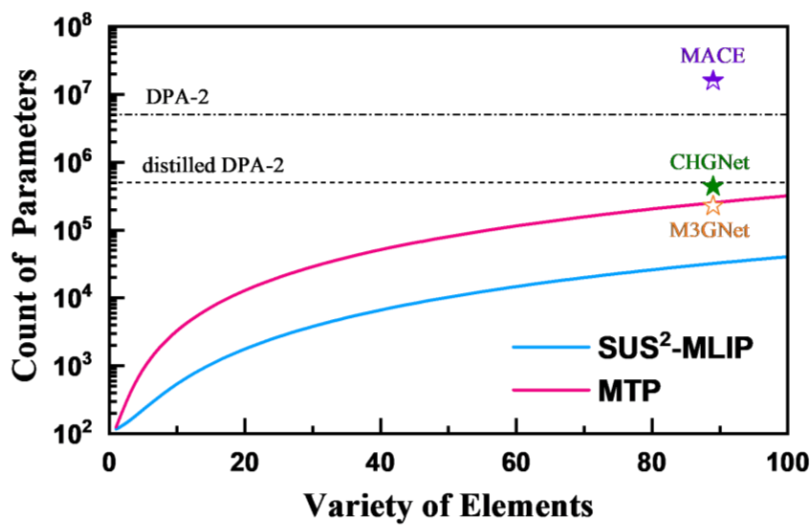

**Fig. 4 Parameter space vs. elemental variety.** The curve of "SUS$^2$-MLIP" is obtained from the specific model labelled as "*l2k2_k*". More details about model labels can be found in *Supplementary Information S.2*. The curve of "MTP" is derived from the model with the level 16 moments. Furthermore, the counts of parameters respective to some neural network models, such as MACE[30], DPA-2[40], CHGNet[8], and M3GNet[25], are also exhibited.

## III.2 Phonons and thermal transport in half-Heusler materials

In semiconductor crystalline materials, the thermal conductivity is governed by phonon transport. The phonon spectrum depends on the curvature of EPS, while their lifetimes are determined by higher derivatives of EPS. Essentially, the fine structures of EPS are not considered in the training of MLIP, which poses challenges to the model's physical scalability as simulating nonlinear behaviors.

Here, we trained an all-in-one $SUS^2$-MLIP model comprising 130 half-Heusler compounds[42] (see Fig. 5(a)) with 54 elements, and studied their phonon transport properties in the framework of Boltzmann transport equation (BTE). As depicted in Fig. 5(b), our model accurately predicts the phonon frequency of all 130 compounds, showcasing a mean relative error (MRE) of 1.91%. Subsequently, we directly derived third-order force constants from MLIP-EPS to compute the lattice thermal conductivity ($\kappa_l$). The calculation details and results of both the phonon spectrum and temperature dependent $\kappa_l$ are detailed in *Supplementary Information S.4*. In Fig. 5(c), we compared our model predictions with DFT calculations for room temperature $\kappa_l$ of 14 half-Heusler materials[43]. The final MRE of $\kappa_l$ is 6.44%, with a maximum absolute error of 3.78 W/mK observed in HfCoSb, demonstrating that $SUS^2$-MLIP can accurately describe nonlinear effects by training solely on energy, forces, and stresses.

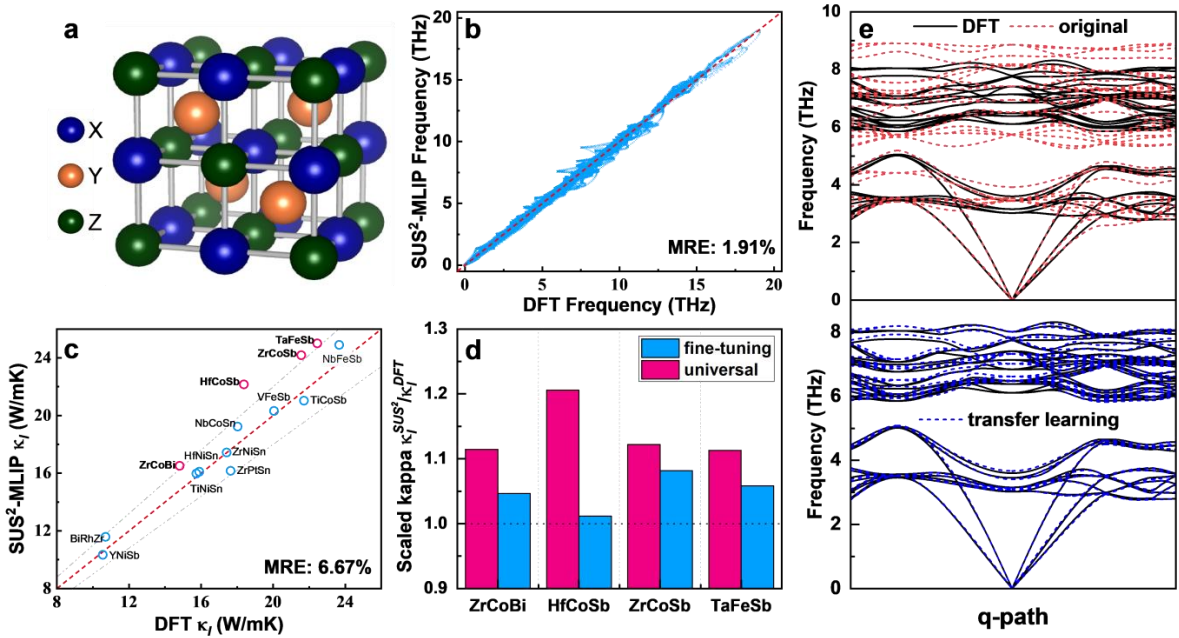

**Fig. 5 The $SUS^2$-MLIP phonons and thermal transports of half-Heusler materials. (a)** The structure of half-Heusler compounds. X and Y are transition metals with weak and strong electronegativities, respectively, and Z is a main group element. **(b)** The comparison of phonon frequency of SUS$^2$-MLIP vs. DTF calculations originated from ref[42]. **(c)** The SUS$^2$-MLIP predicted room temperature $\kappa_l$ for 14 half-Heusler materials against corresponding DTF results derived from ref[43]. **(d)** Further enhancing models' prediction accuracy through finetuning on specific systems. Performances between universal and specific finetuning models on $\kappa_l$ prediction. **(e)** Phonon dispersion of FeCo$_3$Sb$_4$Ti$_4$ before and after transfer learning with small datasets.

The simulation results above are derived from a universal model, which can be further enhanced through targeted finetuning for specific systems. To this end, we selected four systems with an error exceeding 10% as case studies and conducted model finetuning on each system. As indicated in Table-S3, the adjusted model demonstrates improved performance on Energy-Force-Stress. The improvement of phonon transport simulations is evident in Fig. 5(d). Leveraging the universal radial function, SUS$^2$-MLIP can effectively perform transfer learning with limited samples. For instance, in the case study on the solid solution system FeCo$_3$Sb$_4$Ti$_4$ [43], the absence of Co-Fe information results in a significant deviation of the predicted phonon dispersion. After incorporating only 45 additional samples of FeCo$_3$Sb$_4$Ti$_4$, model effectively captures the Co and Fe interaction, leading to a substantial improvement in phonon spectrum predictions (refer to Fig. 5(e)).

**III.3 Molecular dynamics simulations of thermal transport of liquid-like structures**

The integration of diverse components via hierarchical chemical bonding leads to the complex properties of hybrid materials which decorates high-symmetry lattices with low-symmetry groups to achieve an intermediate state that is neither crystalline nor amorphous. The presence of inherent dynamical disorder makes it challenging to accurately describe thermal transport phenomena using perturbation-theory-based phonon models. Therefore, simulating thermal transport in such systems characterized by strong anharmonicity and pronounced disorder poses a considerable challenge for MLIP.

To showcase the physical scalability of our approach, we utilized the SUS$^2$-MLIP driven equilibrium molecular dynamics method to simulate the thermal transport properties of β-Cu$_2$Se[44], in which Cu ions exhibit liquid-like behaviors and diffuse within the rigid Se sublattice. The dynamic disorder of Cu can effectively scatter lattice vibrations and introduce complex transport mechanism. The self-diffusion coefficients from the fits of Stocks-Einstein relation are presented in Fig. 6(a). In the temperature range 500-800 K, Cu diffusivity reaches the order of $10^{-5}$ cm$^2$/s, which is comparable with previous computation studies[45,46]. Due to this remarkable atomic diffusion, the contribution of thermal convection to the whole thermal transport is non-negligible. The temperature-dependent $\kappa_l$ predicted by SUS$^2$-MLIP, as shown in Fig. 6(b), closely corresponds to the experimental measurement[44] which exhibits the temperature-insensitive trend. Besides, we observed that convection's contribution increased with temperature, rising from 0.055 W/mK at 500K to 0.085W/mK at 800K, aligning with the variation in Cu ion diffusion coefficients. The simulation details can be found in *Supplementary Information S.5*.

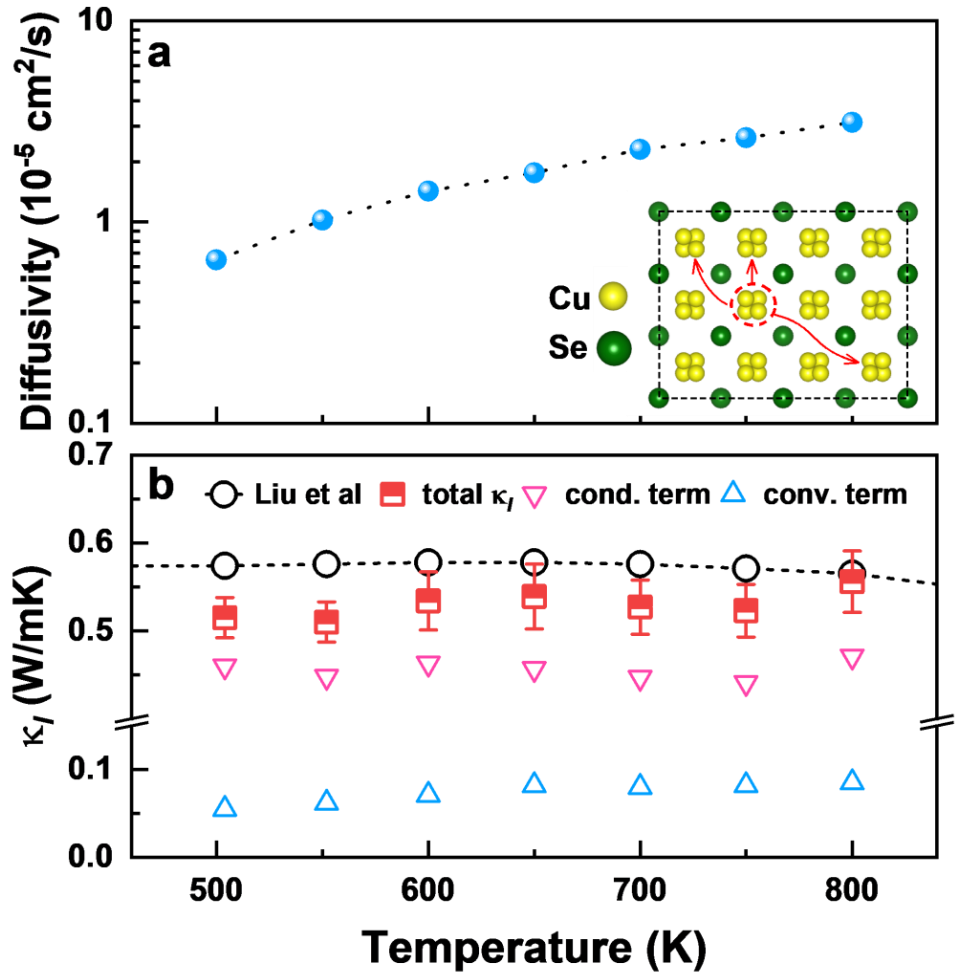

**Fig. 6 SUS²-MLIP predicted transport properties in partial-crystalline-partial-liquid β-Cu₂Se.** Temperature dependences of **(a)** Cu diffusivity and **(b)** $\kappa_l$ contributed from thermal conduction and thermal convection. The error bars represent the standard deviation of multiple simulations. The experimental measurements are derived from the study conducted by Liu et al[44].

**III.4 Li-ion diffusion in Sulfide Solid-state Electrolytes**

To showcase the performance of SUS$^2$-MLIP in predicting kinetic properties and its exceptional simulation efficiency, we employ it to investigate Li-ion migration in LGPS-like[47] Li$_{10}$XP$_2$S$_{12}$ (X=Ge, Si, Sn) system that have garnered significant attention due to their high ionic conductivity and remarkable chemical stability. The structure of Li$_{10}$XP$_2$S$_{12}$ is shown in Fig. 7(a). Here, we trained a SUS$^2$-MLIP model using the dataset of sulfide solid-state electrolytes from ref[48], which comprises 54,771 structures and encompasses 15 elements. The Li diffusivity and activation energy predicted by SUS$^2$-MLIP in Fig. 7(b) agrees with both DPA-2 predictions and *ab initio* molecular dynamics (AIMD) results, demonstrating our approach precisely captures global diffusion properties under various thermodynamic conditions. More simulation details can be found in *Supplementary Information S.6*.

In the field of materials science, numerous distinctive phenomena require simulations involving a large number of atoms and extended timescales. To address this, we evaluated the computational efficiency of both SUS$^2$-MLIP and DPA-2 across various system sizes of Li$_{10}$GeP$_2$S$_{12}$. During the scaling test, computational resources were held constant while the number of atoms was varied. Fig. 7(c) clearly demonstrate that the SUS$^2$-MLIP exhibits a significantly higher efficiency, approximately 5-6 times greater than that of DPA-2 when implemented on a high-performance GPU device (NVIDIA-A100). Notably, due to memory constraints, DPA-2 can only simulate a maximum of about 70,000 atoms. In contrast, the resource-efficient SUS$^2$-MLIP demonstrates remarkable scalability and allows for the simulation of large-scale systems exceeding 1 million atoms. Finally, we evaluate the parallel computing efficiency of SUS$^2$-MLIP. Fig. 7(d) shows the scaling results on Li$_{10}$GeP$_2$S$_{12}$ structure with 72600 atoms. The system size was kept constant while varying the number of computing nodes. When 5 nodes are used to increase the speed of simulation, the actual computational efficiency is decreased by 9.2% compared with the ideal case.

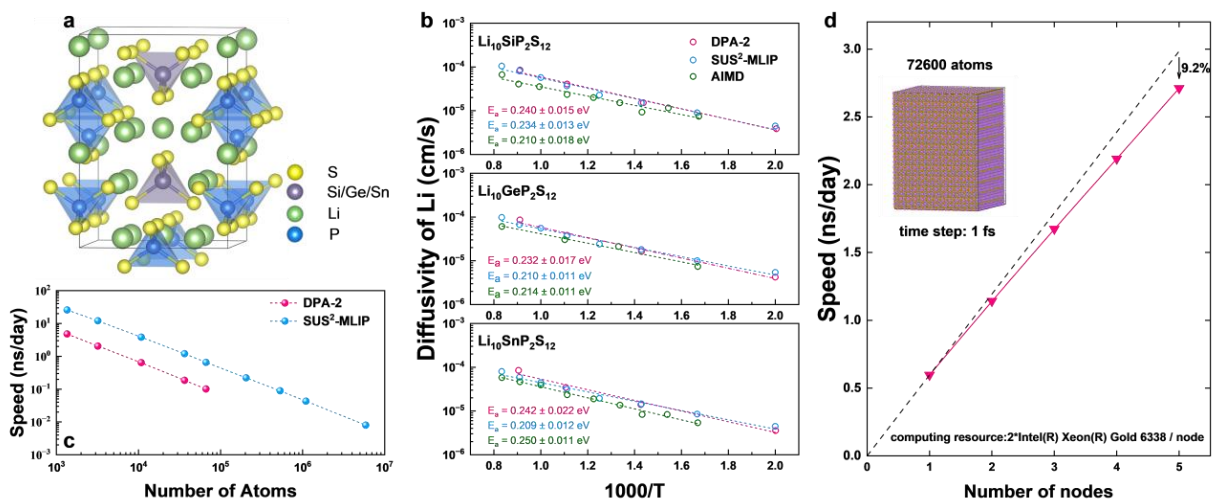

**Fig. 7 Simulation experiments on Li-ion diffusion in sulfide solid-state electrolytes. (a)** Schematic diagram of $Li_{10}XP_2S_{12}$ (X=Ge, Si, Sn) system containing 50 atoms. **(b)** Diffusivity of Li ions at different temperatures and activation energy ($E_a$). The results of AIMD and DPA-2 are derived from ref [48]. **(c)** Scaling test on $SUS^2$-MLIP and DPA-2 models. In the test, the DPA-2 model was performed on 1*NVIDIA-A100 with 80GB video memory and the $SUS^2$-MLIP model was performed on 2*Intel(R) Xeon(R) Gold 6338 CPU with 512 GB memory. When the number of simulated atoms reaches 72600 (corresponding to supercell 12×11×11), the NVIDIA-A100 is out of memory. **(d)** Parallel computing efficiency of $SUS^2$-MLIP.

# IV. Discussions and Conclusions

While numerous studies have demonstrated that ML techniques offer a cost-effective and efficient approach for constructing precise EPS of materials, it is important to note that ML, as a purely mathematical tool, can only yield potential mathematical expressions based on limited datasets but lacks the capacity to provide any physical insights. The gap between "mathematical solutions" and "physical solutions" underlies the challenges of model generalizability and physical scalability. Therefore, the profound comprehension of the fundamental physics that underlies the MLIP model is imperative for its meticulous construction. Besides, an increased prior understanding of the physical image enables a simplification of modeling complexities. The essence of parsimonious modeling lies not only in enhancing computational efficiency but also in abstracting complex systems into forms controlled by the fewest key parameters, thereby highlighting the main physical mechanisms. In fluid dynamics, for example, the behavior of incompressible flows is entirely characterized by the Navier-Stokes equations and the dimensionless Reynolds number derived from global scaling. Analogously, within the $SUS^2$-MLIP framework, all two-body interactions are uniquely determined by a set of universal radial functions and element-dependent scaling factors ($\alpha_{Z_I Z_j}$ and $r_0^{Z_I Z_j}$). The incorporation of global scaling not only drastically reduces the dimensionality of the parameter space but also enhances the model's capacity to reveal underlying fundamental physical principles. Through benchmark tests and property simulations, we demonstrate that $SUS^2$-MLIP, despite its lightweight architecture, achieves a comparable level of accuracy to more complex models. Additionally, it exhibits superior computational efficiency and the capability to perform large-scale simulations involving millions of atoms.

While $SUS^2$-MLIP has yielded the aforementioned outcomes, we contend that a crucial objective for future research is to construct MLIP models beyond mean-field approximations by implementing parsimonious modeling techniques. For instance, one could elucidate indirect interactions by integrating appropriate neural network architectures upon generalized linear models, or directly decompose fully coupled multi-body interactions into one-dimensional functions via the KAN framework to achieve model simplification. Besides, we believe the quantitative characterization of sample space completeness, derived from atomic environmental descriptor, constitutes a significant and promising research direction for future exploration.

In summary, we developed SUS$^2$-MLIP, an ultracompact MLIP model derived from a many-atomic field, which owns the comparable accuracy with complex models, outstanding physical scalability and ability to scale to large system simulations. We expect this method can not only enable simulations of complex systems in various research domains of materials science, but also open new horizons for the design of physics-informed MLIP models.

# Methods

**Reference training sets**

**Alloy**[*URL]. The *alloy* dataset comprises approximately 150,000 structures and encompasses 53 metallic elements, including random substitutional solid solutions, elemental substances, and intermetallic compounds. We trained the SUS$^2$-MLIP model using the "*l2k3_k*" basis with the radial cutoff of 6.0 Å, which consists of 17,689 parameters.

**Cathode**[*URL]. The *cathode* dataset investigates layered transition metal oxide cathodes (Li/Na)$_x$-TMO$_2$ with TM $\in$ {Ni, Mn, Fe, Co, Cr} and x $\in$ {0, 0.5, 1}. The original training set contains 88,692 structures, 1/3 of which (29564) were used for model training on "*l3k3_lk*" basis with 2,066 parameters and the radial cutoff of 6.5 Å.

**Perovskite**[*URL]. The *perovskite* dataset comprises 26 ABO$_3$ perovskite oxides spanning from 3-element to 6-element, with A $\in$ {Ba, Pb, Ca, Sr, Bi, K, Na} and B $\in$ {Hf, Ti, Zr, Nb, In, Zn, Mg}. We used the "*l3k3_lk*" basis to train our model with a total budget of 6,966 structures and considered the radial cutoff of 6.5 Å. This modeling consists of 6,028 parameters.

**Semiconductors**[*URL]. The *semiconductor* dataset investigates 20 commonly used semiconductors under various thermodynamic conditions, which contains 215,481 frames for training and 23,343 for validation. In this work, the training (101,967) and validation (11,183) datasets were chosen based on a criterion of max atomic force < 5 eV/Å. We used 20,394 configurations for model training on "*l3k3_lk*" basis with 6,786 parameters, and considered the radial cutoff of 6.0 Å.

**MPF.2021.2.8**[*URL]. The dataset of M3GNet comprises 187,687 structures and encompasses 89 elements. We trained the SUS$^2$-MLIP model using the "*l2k2_k*" basis with the radial cutoff of 6.0 Å, which consists of 29,006 parameters.

**Ion diffusion simulations**

According to the Stocks-Einstein relation, the self-diffusion coefficient $D$ can be calculated as:

$$D = \lim_{t \to \infty} \frac{MSD(t)}{6t} \tag{9}$$

where, the mean square displacement $MSD(t) = \frac{1}{N}\sum_i^N |r_i(t) - r_i(0)|^2$ can be derived from the trajectory of a sufficiently long molecular dynamics simulation. The diffusion coefficient $D$ and temperature $T$ satisfy the Arrhenius relationship:

$$D(T) = D_0 \exp\left(-\frac{E_a}{k_B T}\right) \tag{10}$$

where the activation energy $E_a$ can be obtained by fitting a linear relationship of $\ln D \sim \frac{1}{T}$.

**Phonon transport simulations**

Phonon thermal conductivity tensor element of αβ direction at temperature T can be calculated as the sum of contributions from each phonon modes λ:

$$\kappa_{\alpha\beta} = \frac{1}{NV} \sum_\lambda \frac{\partial f_\lambda}{\partial T}(\hbar\omega_\lambda) v_\lambda^\alpha v_\lambda^\beta \tau_\lambda \tag{11}$$

where, $N, V, f_\lambda$ and $v$ is the the number of wave vector q points in the Brillouin zone, the volume of unit cell, the phonon frequency $\omega_\lambda$ dependent equilibrium Boltzmann distribution and phonon group velocity, respectively.

**Thermal conduction simulation based on Green-Kubo theory**

Based on Green-Kubo theory, the thermal conductivity can be expressed as:

$$\kappa_l(\tau) = \frac{1}{k_B T^2 \Omega} \int_0^\tau <J(0)J(t)> dt \tag{12}$$

where $k_B$ is the Boltzmann constant and $\Omega$ the volume of simulation box. The angle brackets $<>$ denotes the correlation function and $J(t) \equiv \frac{d}{dt}\sum_i r_i(t) U_i(t)$ is the transient heat flux at time $t$. $m_i$ and $v_i$ refer to the mass and velocity of atom i, respectively.

**99**, 014104 (2019).

**Acknowledgements:** This project is supported by the National Natural Science Foundation of China (92163212, 92463310), National Key R&D Program of China (2022YFA1203400), High Level of Special Funds (G03050K002) and Guangdong Provincial Key Laboratory of Computational Science and Material Design (2019B030301001). Computing resources were supported by the Center for Computational Science and Engineering at Southern University of Science and Technology.

**Author contributions:** Y.H., Y.S. and W.Z. conceived the initial idea. J.H., X.X. and Y.Y. collected the datasets. Y.H. developed and formalized the code base. Y.H., J.H. and M.Z. performed the simulations and analyzed the data. W.Z., Y.S., C.Y., Y.W. and J.Y. offered insight and guidance throughout the project. Y.H. and W.Z wrote the manuscript and contributed to the discussion and revision.

**Code availability:** An open-source software implementation of SUS$^2$-MLIP is available at https://github.com/hu-yanxiao/SUS2-MLIP/

**Competing interests:** The authors declare no competing interests.

# Supplementary Information

**Universal Scaling and Ultra-Small Parameterization in Machine Learning Interatomic Potentials with Super-Linearity**

**Authors:** Yanxiao Hu [1†], Ye Sheng [1†], Jing Huang [1], Xiaoxin Xu [1,2], Yuyan Yang [2], Mingqiang Zhang [1], Yabei Wu [1,3], Caichao Ye [1,3], Jiong Yang [2] and Wenqing Zhang [1,3*]

**Affiliations:**

[1] Department of Materials Science and Engineering, Southern University of Science and Technology, Shenzhen, Guangdong 518055, China

[2] Materials Genome Institute, Shanghai Engineering Research Center for Integrated Circuits and Advanced Display Materials, Shanghai University, Shanghai 200444, China.

[3] Institute of Innovative Materials & Guangdong Provincial Key Lab for Computational Science and Materials Design, Southern University of Science and Technology, Shenzhen, Guangdong 518055, China

*Corresponding author. Email: zhangwq@sustech.edu.cn

†These authors contributed equally to this work.

## S1. Mean field approximation

In many-atomic systems with complex interdependencies, as shown in Fig.S1, the interaction from environmental atom $j$ to atom $I$, $\Phi_{Ij}$, can be expressed as the contributions from each path:

$$\Phi_{Ij} = \varphi_{Ij} + \underbrace{\sum_{k} \varphi_{Ij}^{k} + \sum_{k,l} \varphi_{Ij}^{kl} + \cdots}_{\widetilde{\varphi}_{Ij}(n_j)} \tag{S1}$$

Here, $\varphi_{Ij}$ presents the direct interaction that only depends on the pair vector $\boldsymbol{r}_{Ij} \equiv \boldsymbol{r}_j - \boldsymbol{r}_I$ as well as corresponding elemental type $Z_I$ and $Z_j$. While, the remaining terms represent the indirect interaction resulting from the atomic environment $n_j$. For non-dense systems, wherein direct interactions are predominant, it is reasonable to approximate indirect terms as a constant mean field, namely $\Phi_{Ij} \approx \varphi_{Ij} + 1$.

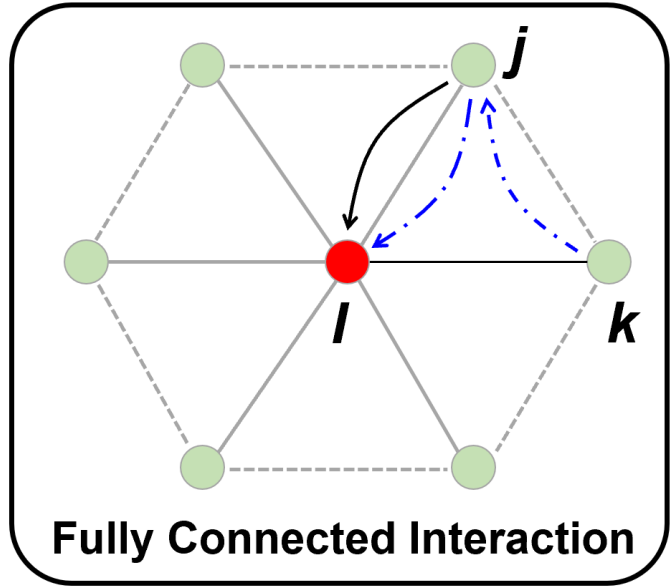

**Fig. S1 Real many-atomic system.** Atoms are coupled together through complex interactions, making it difficult to analyze the behavior of individual atoms.

## S2. Construction of SUS²-MLIP

**Universal radial function**

The basic relationship between elemental pair $r_{Ij}$ and scaled pair $r_{Ij}^*$ is expressed as: $r_{Ij}^* = \alpha_{Z_I Z_j}(r_{Ij} - r_0^{Z_I Z_j})$. Considering the scaling factors not only dependent on element but also on specific crystal structure and representation space, we incorporated an additional channel $\eta$ in scaling process, namely, $r_{Ij,\eta}^* = \alpha_{Z_I Z_j,\eta}(r_{Ij} - r_0^{Z_I Z_j,\eta})$. In our program, we accommodate three distinct scaling modes, regarding $kl$, $k$ and $l$, which corresponds to $\eta = (k,l)$, $\eta = k$ and $\eta = l$ respectively. Then, the radial function can be written as:

$$\tilde{R}_l^{(k)} = \left(\sum_n C_n^{l,k} T_n(x_{Ij}^\eta)\right) * f_{cutoff} \tag{S2}$$

where $T_n$ is the $n$-th order of Chebyshev polynomials and $C_n^{l,k}$ the expansion coefficient. To ensure the smooth behavior of radial function, the cutoff function was introduced:

$$f_{cutoff} = \begin{cases} (R_{cut} - |r_{Ij}|)^2 & (|r_{Ij}| < R_{cut}) \\ 0 & (|r_{Ij}| > R_{cut}) \end{cases} \tag{S3}$$

The $x_{Ij}^\eta$, ranging from -1 to 1, is obtained from $r_{Ij,\eta}^*$ according to the transformation:

$$x_{Ij}^\eta = \tanh(r_{Ij,\eta}^*) \tag{S4}$$

We believe *tanh* is a good choice because it is in line with the fundamental atomic interaction behavior that the nonlinear effect becomes increasingly pronounced as the atom moves further away from its equilibrium position.

**Descriptors of many-body interactions**

In this work, the moment tensors of directional vector are employed to characterize the directional information of atomic pairs. The $l$-th order of moment tensor $\hat{r}^{\otimes l}$ has a dimensionality of $[\overbrace{3 \times ... \times 3}^{l}]$, which also respective to a representation space of *SO(3)* group. Then, the $l$-th subspace of basis function for single pair $r_{Ij}$ can be expressed as:

$$\boldsymbol{B}_l^{(k)}(\vec{r}_{Ij}) = \tilde{R}_l^{(k)}(r_{Ij,\eta}^*)\hat{r}_{Ij}^{\otimes l} \tag{S5}$$

For example, if $l=0,1$ and 2:

$$\boldsymbol{B}_0^{(k)}(\vec{r}_{Ij}) = \tilde{R}_0^{(k)}$$
$$\boldsymbol{B}_1^{(k)}(\vec{r}_{Ij}) = \tilde{R}_1^{(k)}[\hat{x}_{Ij}, \hat{y}_{Ij}, \hat{z}_{Ij}]$$
$$\boldsymbol{B}_2^{(k)}(\vec{r}_{Ij}) = \tilde{R}_2^{(k)} \begin{bmatrix} \hat{x}_{Ij}^2 & \hat{x}_{Ij}\hat{y}_{Ij} & \hat{x}_{Ij}\hat{z}_{Ij} \\ & \hat{y}_{Ij}^2 & \hat{y}_{Ij}\hat{z}_{Ij} \\ & & \hat{z}_{Ij}^2 \end{bmatrix}$$

By collecting the contribution of each pair, the atomic-centered *2*-body interactions $\{\boldsymbol{B}^{l,k}(n_I) = \sum_j \boldsymbol{B}_l^{(k)}(\vec{r}_{Ij})\}$ are obtained. While the $n+1$-body interactions can be naturally expressed as direct product of 2-body subspace: $\boldsymbol{B}^{l_1...l_n,k_1...k_n} = \boldsymbol{B}^{l_1,k_1} \otimes ... \otimes \boldsymbol{B}^{l_n,k_n}$.

Scalar descriptors $\{D_\alpha\}$ can be extracted from many-body interaction space by specific inner product process $\alpha$. For the simplest *2*-body interactions, scalar subspace is directly obtained as $l=0$, namely, $D_{\alpha^{(2)}=(0,k)} = B^{0,k}$, and $<\varphi_I>_0 = \sum_{\alpha^{(2)}} C_{\alpha^{(2)}} D_{\alpha^{(2)}}$. As to $n$-body interactions, we provide two examples for clarification. The inner product process of 3-body term $\alpha^{(3)} = (1\ 1,\ k_1 k_2)$ refers to $D_{\alpha^{(3)}=(1\ 1,k_1 k_2)} = \sum_{\beta \in (x,y,z)} B_\beta^{1,k_1} B_\beta^{1,k_2}$, while the 4-body term $\alpha^{(4)} = (2\ 1\ 1,\ k_1 k_2 k_3)$ refers to the $D_{\alpha^{(4)}=(2\ 1\ 1,k_1 k_2 k_3)} = \sum_{\beta_1,\beta_2} B_{\beta_1 \beta_2}^{2,k_1} B_{\beta_1}^{1,k_2} B_{\beta_2}^{1,k_3}$. Finally, the $<\varphi_I^n>_0$ can be expressed as the following linear combination:

$$<\varphi_I^n>_0 = \sum_{\alpha^{(2)}} C_{\alpha^{(n)}} D_{\alpha^{(n)}} \tag{S6}$$

where $C_{\alpha^{(n)}}$ is the learning expansion parameter.

In this work, we have used two sets of models corresponding to the $l_{max} = 2 \ or \ 3$, in both of which the interactions are considered up to the 5-body. More details about scalar basis in each model are listed in Table-S1. The max level of $k$ channels is considered to 3, hence, we prepared 6 sets of untrained basis: *l2k1_η, l2k2_η, l2k3_η, l3k1_η, l3k2_η* and *l3k3_η*.

**Table-S1. Specific inner product processes in models implemented in this work.**

| Models | | Body Order | | | |
|---|---|---|---|---|---|
| | | 2 | 3 | 4 | 5 |
| $l_{max} = 2$ | <0> | <00>,<11>,<22> | <000>,<011>,<022>,<112>,<222> | <11><11>,<11><22>,<22><22> |
| $l_{max} = 3$ | <0> | <00>,<11>,<22>,<33> | <000>,<011>,<022>,<033>,<112>,<123>,<222>,<233> | <11><11>,<11><22>,<11><33>,<22><22>,<22><33>,<33><33> |

**The <> denotes a specific inner product procedure irrespective of the channel $k$. For example, <11> is respective to 3-body term $\alpha^{(3)} = (1\ 1,\ k_1 k_2)$. To simplify the construction of high-order descriptor, all 5-body terms are derived from the products of 3-body terms.**

**Model regression**

The SUS²-MLIP models are trained to minimize the joint loss function of energy, force and stress:

$$Loss = \sum_{y=e,f,s} w_y \mathcal{L}_y \quad (S7)$$

where, the weights for energy, force and stress are initially set to $w_e$=1, $w_f$=0.01 and $w_s$=0.001. The detail definition of $\mathcal{L}_y$ can be found in ref[1]. The line search based BFGS method is used to optimize the model. The initial radial parameters $\{C_n^{l,k}\}$ are generated randomly and normalized. While, the expansion parameters $\{C_{\alpha^{(n)}}\}$ and refence energy $\widetilde{\varepsilon_{Z_I}}$ are initialized using a least squares solution.

**Element-property embedding beyond mean field approximation**

Under the mean field approximation, we consider $t_I$ is solely dependent on corresponding atomic type, which is reasonable for homogenous systems. However, when dealing with systems with significant chemical environment diversity, such as tri- and tetra-coordinated carbon atoms, $t_I$ should be a complex mapping of atomic environment. In crystal graph neural networks, such as CGCNN, $t_I$ is initially presented by a long node vector that consists of atomic features e.g. covalent radius, valence electrons, etc. Through graph convolution, $t_I$ updates and obtains surrounding environmental information. In graph based MLIP models, $t_I$ is simplified to a one-hot vector. This approach has been demonstrated to significantly enhance models' performance, as graph convolution to some extent describes the indirect interactions as discussed in *section S1*, and gives atomic interactions that goes beyond the mean field approximation. Physically, $t_I$ should be expanded to an atomical property field $\mathcal{T}_I$:

$$\mathcal{T}_I(\boldsymbol{r}_I; \{\boldsymbol{r}_j\}) = t_I \left(1 + <\tilde{\varphi}_I>_0 + \frac{1}{2}<\tilde{\varphi}_I^2>_0 + \cdots \right) \quad (S8)$$

When the environmental effect is disregarded, the aforementioned equation degenerates to $t_I$.

## S3. Additional datasets

In addition to benchmark, we also conducted model training on additional datasets to illustrate the application contexts of SUS$^2$-MLIP from simple crystal materials to hierarchical hybrid materials. The final model performances are listed in Table-S2.

**HH130**[*URL]. This dataset investigates 130 half-Heusler compounds, containing 31891 frames and encompassing 54 elements. We trained the SUS$^2$-MLIP model using the "*l2k3_k*" basis with the radial cutoff of 6.0 Å, which consists of 18,310 parameters.

**Cu$_{2-x}$Se**. This in-house dataset comprises 836 configurations, encompassing both high-temperature phases (β-Cu$_2$Se), low-temperature phases (α-Cu$_2$Se), and vacancy-doped structures. We trained the SUS$^2$-MLIP model using the "*l3k2_k*" basis with the radial cutoff of 6.5 Å, which consists of 209 parameters.

**Sulfide Solid-state Electrolytes**[*URL]. This dataset comprises 54,771 structures and encompasses 15 elements. We trained the SUS$^2$-MLIP model using the "*l2k3_lk*" basis with the radial cutoff of 6.0 Å, which consists of 4494 parameters.

**Table-S2 Energy, Force and Stress MAE, in units of [meV/atom], [meV/Å] and [GPa] respectively.**

| Dataset | Energy | Force | Stress |
| --- | --- | --- | --- |
| HH130 | 1.35 | 21.3 | 0.022 |
| Cu$_{2-x}$Se | 1.88 | 31.9 | 0.047 |
| Sulfide-Solid-state Electrolytes | 5.11 | 57.9 | 0.091 |

## S4. Phonon simulations based on HH130 dataset

In this work, both the 2nd and 3rd interatomic force constants of 130 half-Heusler compounds are directly calculated by SUS$^2$-MLIP model with the 3×3×3 supercell. The phonon dispersions and thermal conductivity are obtained through *phonopy* and *phono3py* packages[2,3]. The interaction range are considered to 0.6 nm and the phonon-phonon scattering calculation are implemented on the 31×31×31 q-mesh. Individual results of whole 130 materials are illustrated in Fig. S2&S3.

Finetuning in section *III.2* are conducted by performing second training on dataset excluding data irrelevant to target system. Table-S3 shows the model performance on Energy-Force-Stress before and after finetuning.

**Table-S3 Energy, Force and stress MAE of universal and fine-tuning models, in units of [meV/atom], [meV/Å] and [GPa] respectively.**

| compounds | models | Energy | Force | Stress |
|---|---|---|---|---|
| ZrCoBi | Universal | 1.93 | 35.2 | 0.096 |
|  | Fine-tuning | 1.87 | 21.1 | 0.074 |
| HfCoSb | Universal | 4.05 | 31.2 | 0.086 |
|  | Fine-tuning | 3.91 | 18.8 | 0.065 |
| ZrCoSb | Universal | 0.40 | 30.1 | 0.044 |
|  | Fine-tuning | 0.34 | 19.3 | 0.024 |
| TaFeSb | Universal | 3.81 | 40.3 | 0.113 |
|  | Fine-tuning | 3.62 | 26.6 | 0.081 |

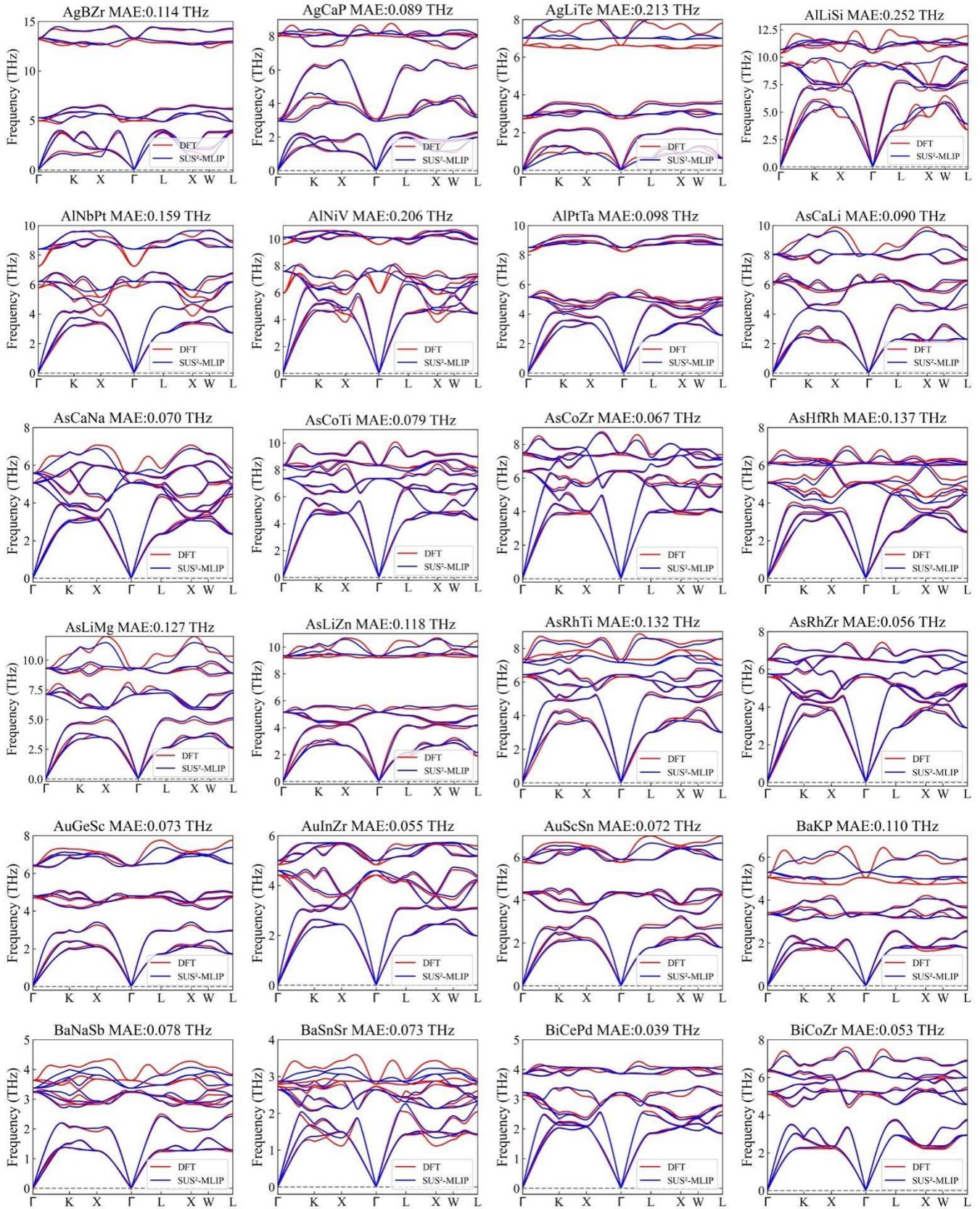

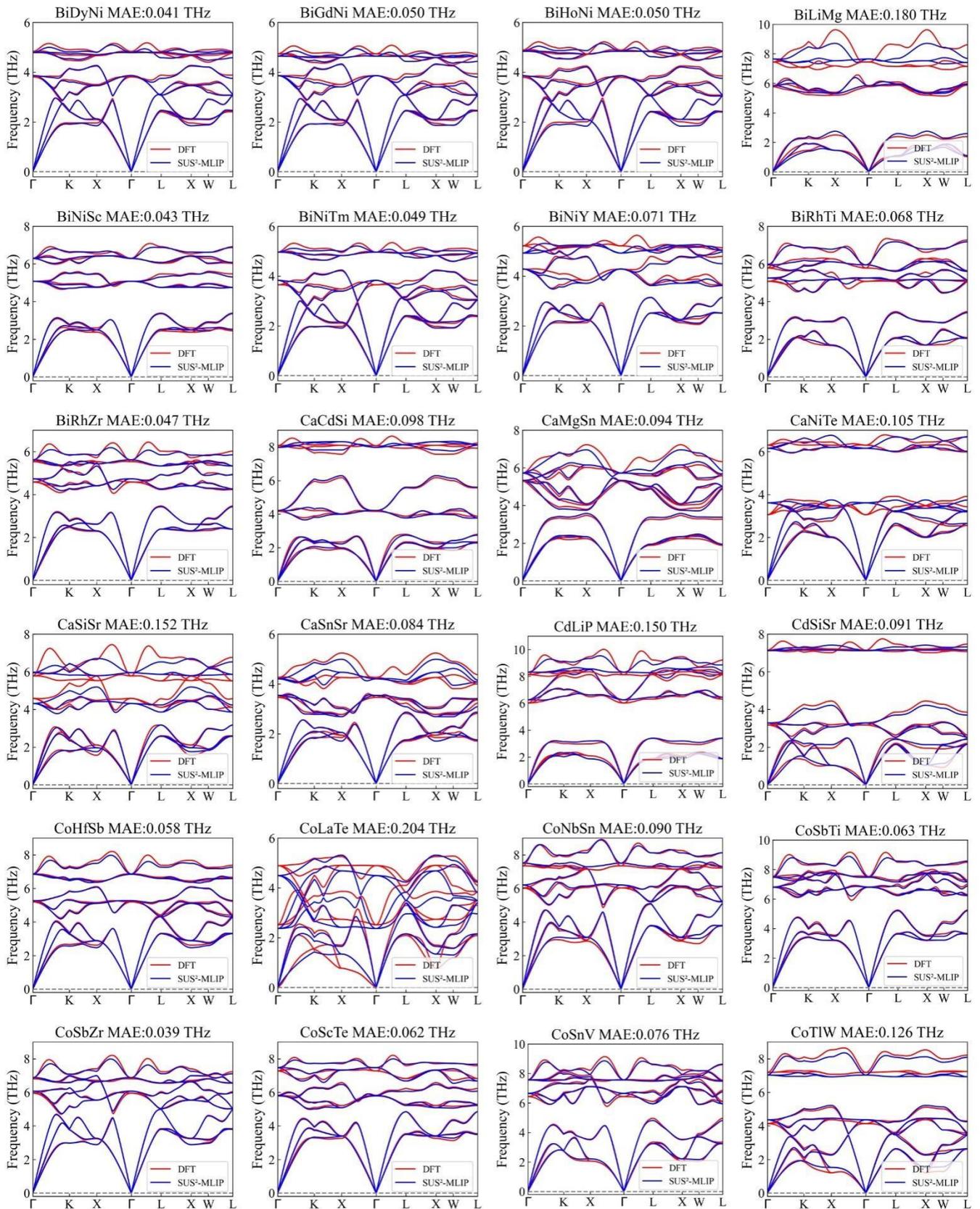

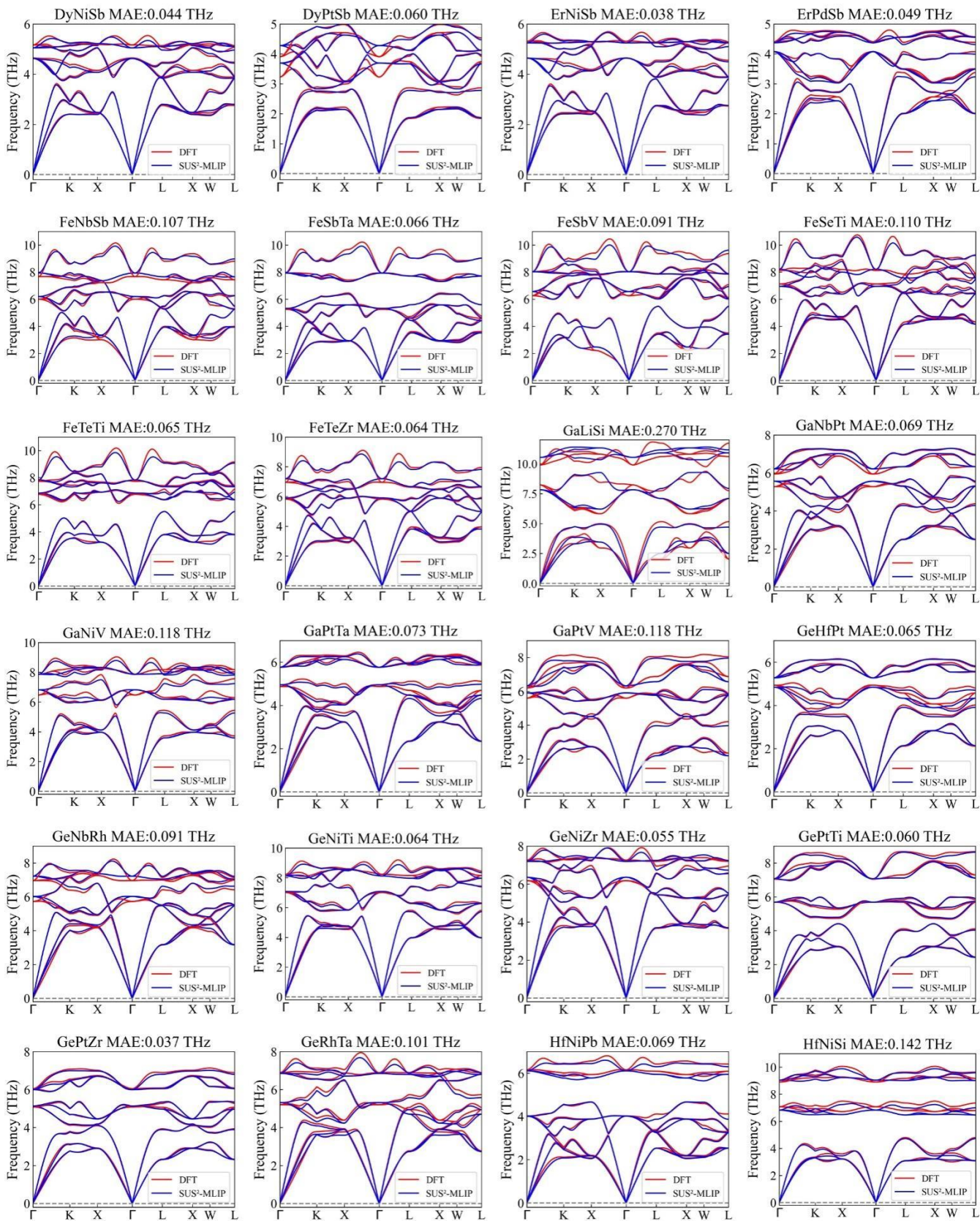

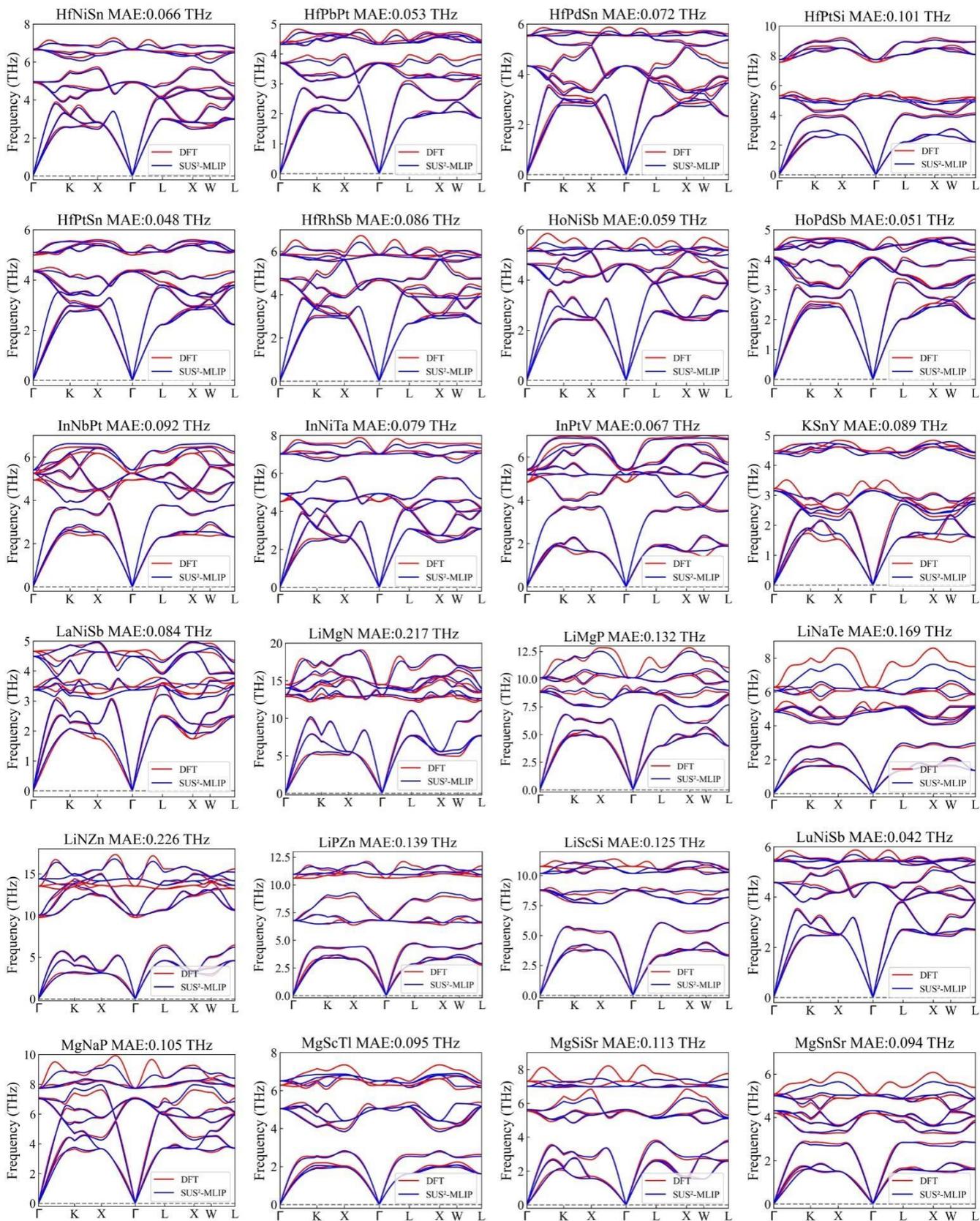

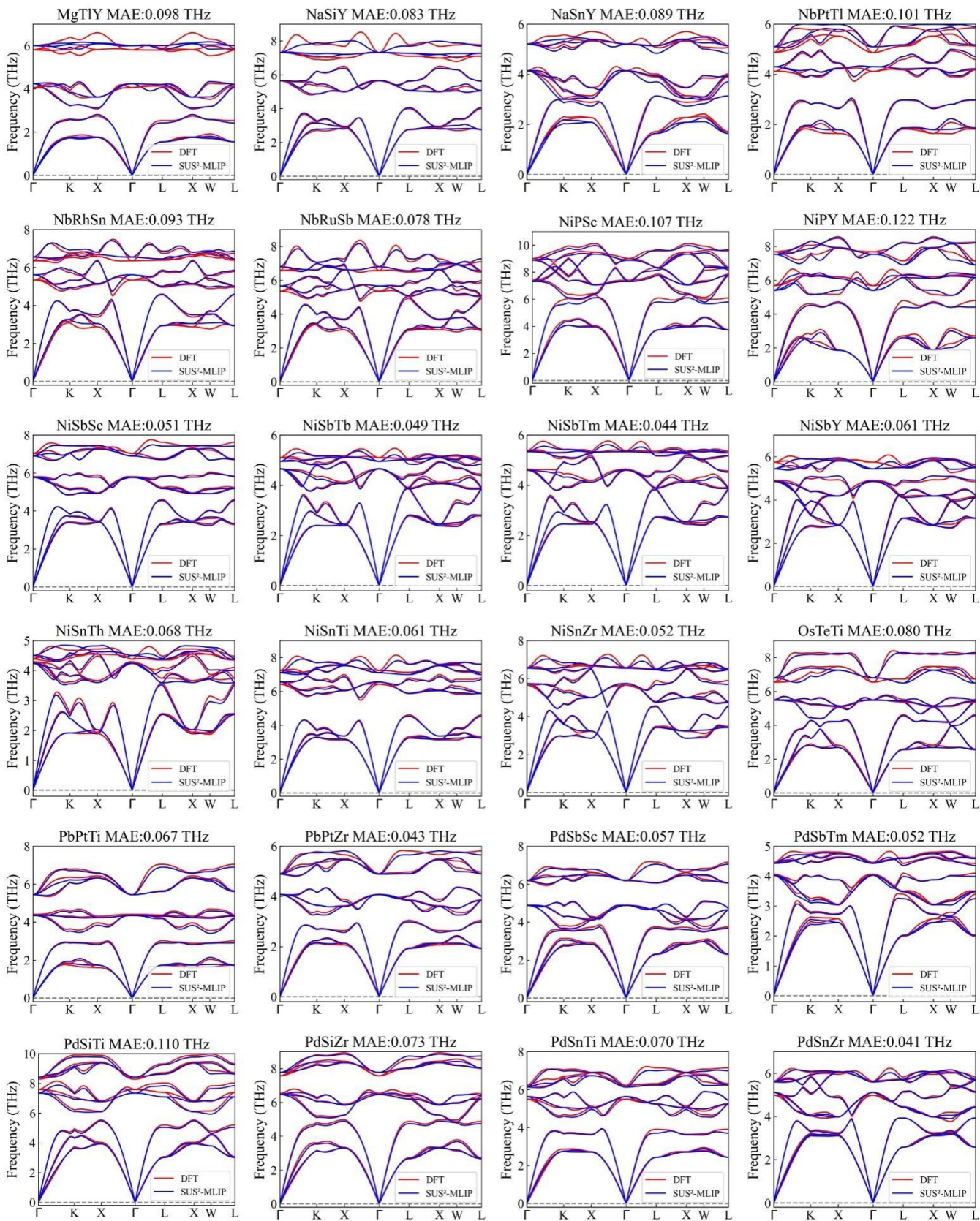

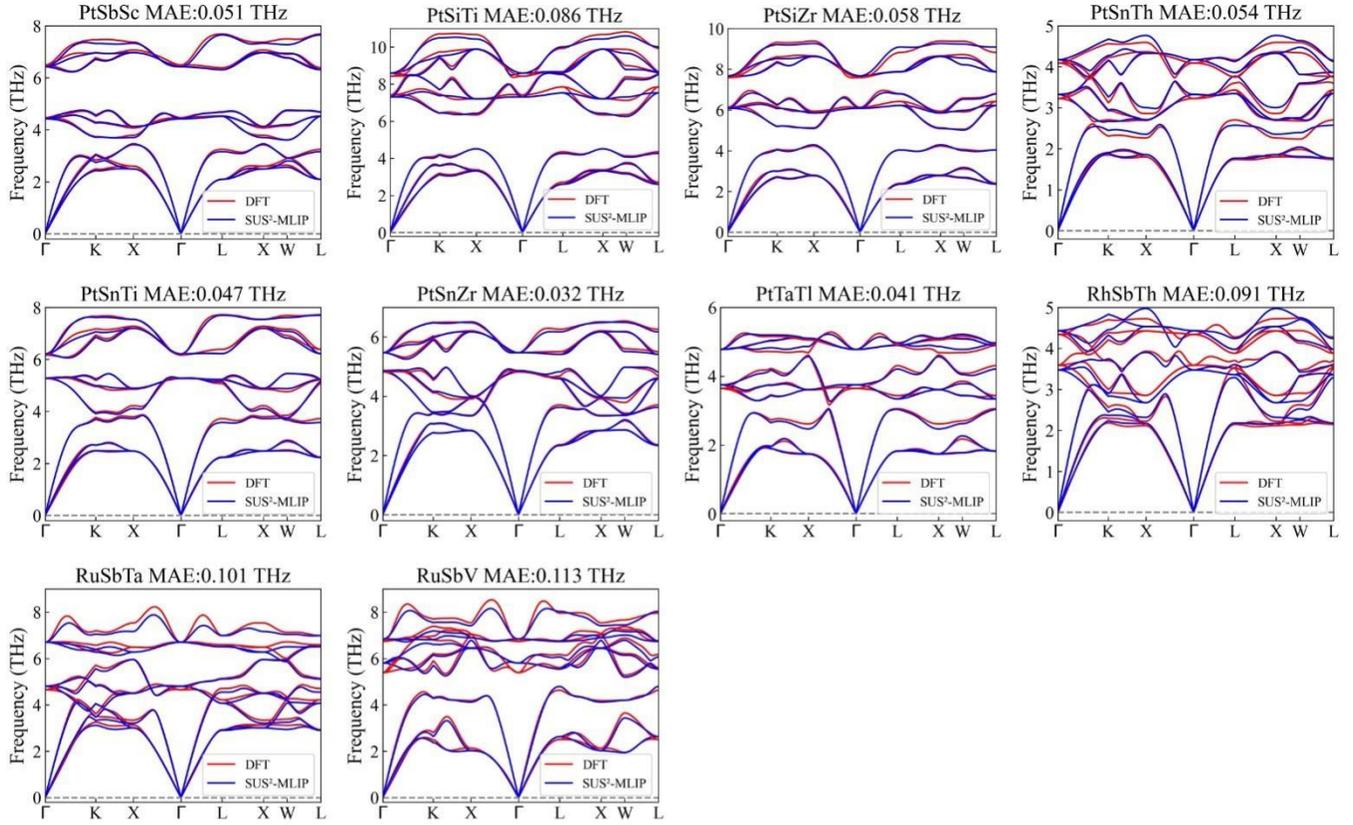

**Fig. S2. Phonon dispersions of 130 half-Heusler compounds.**

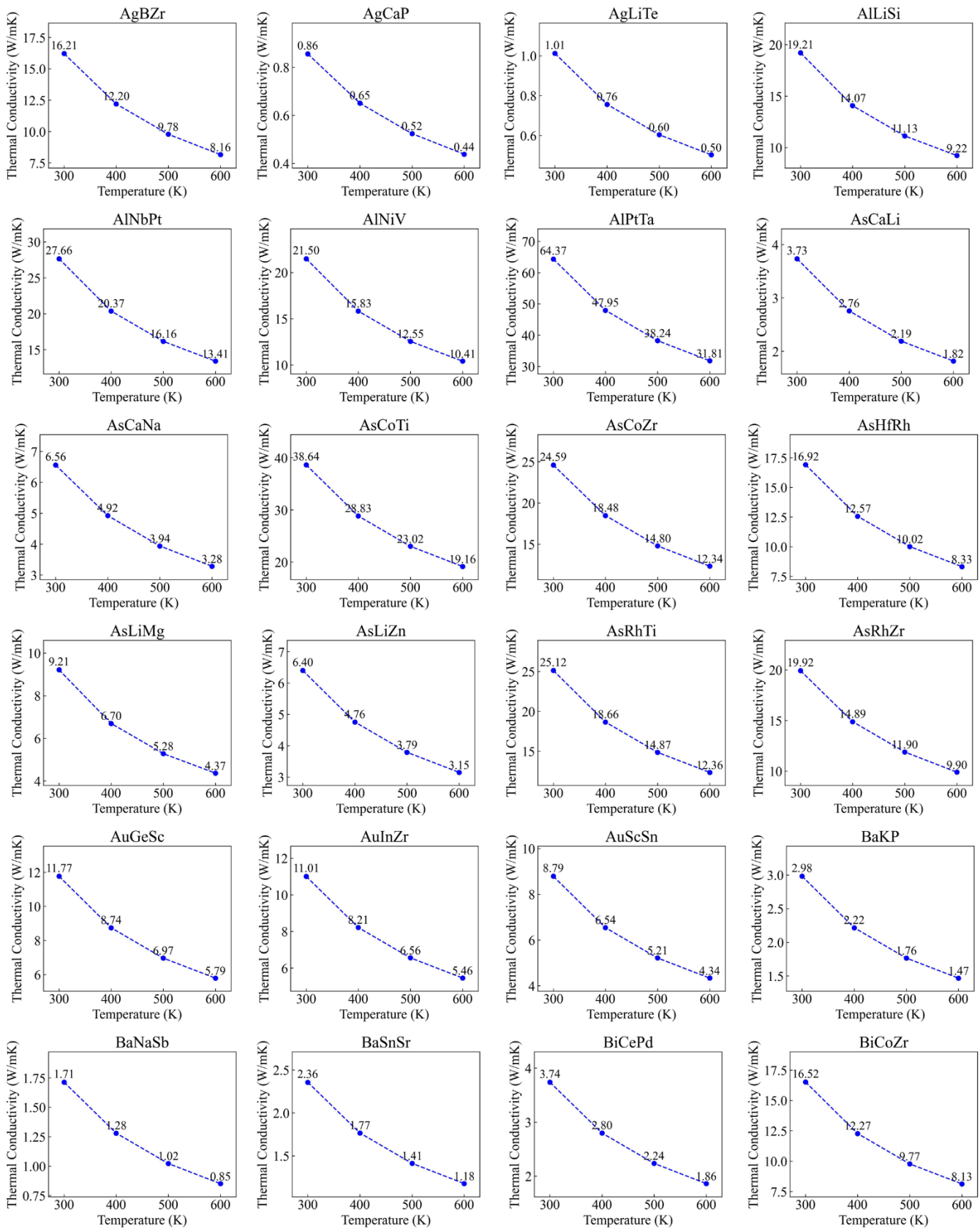

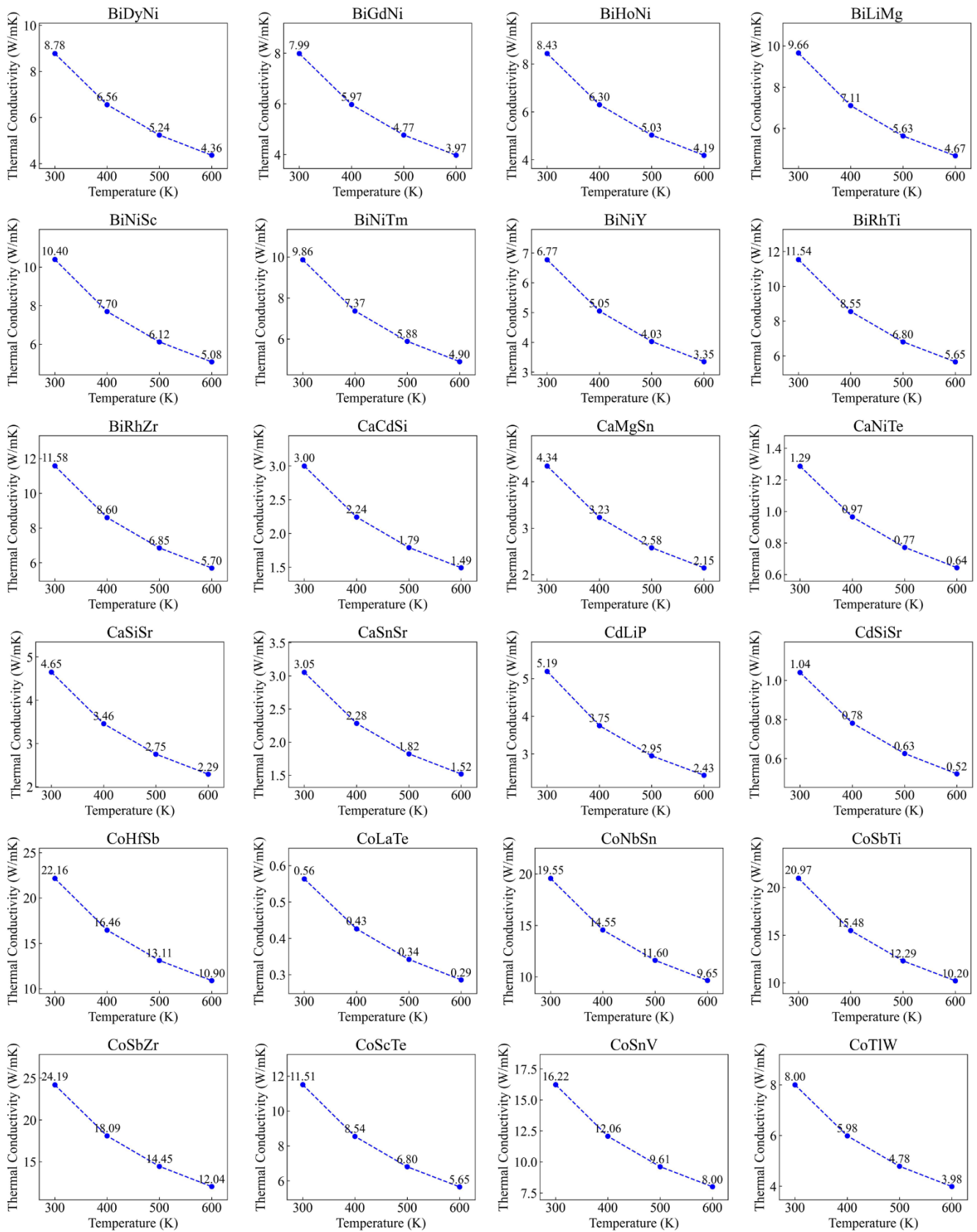

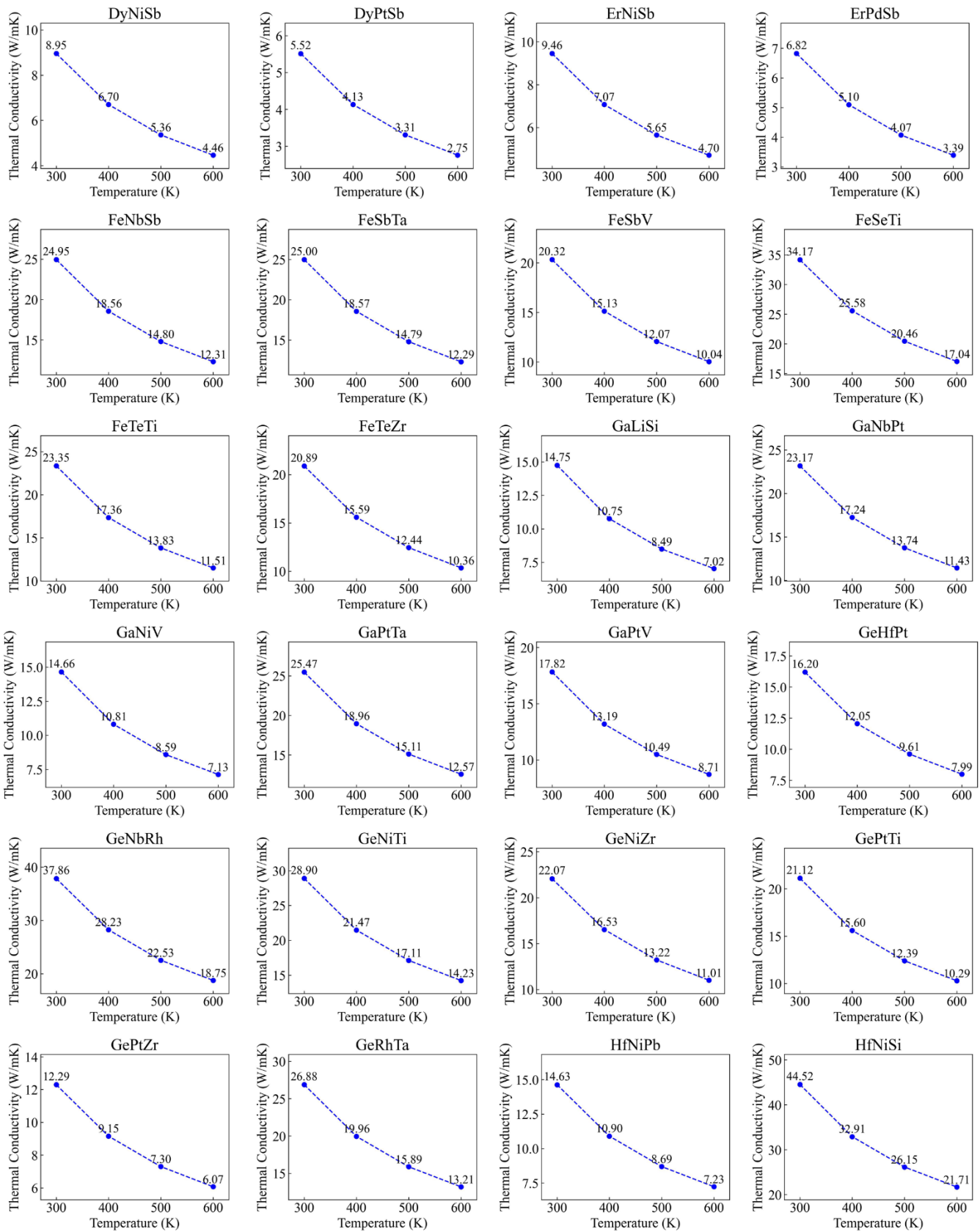

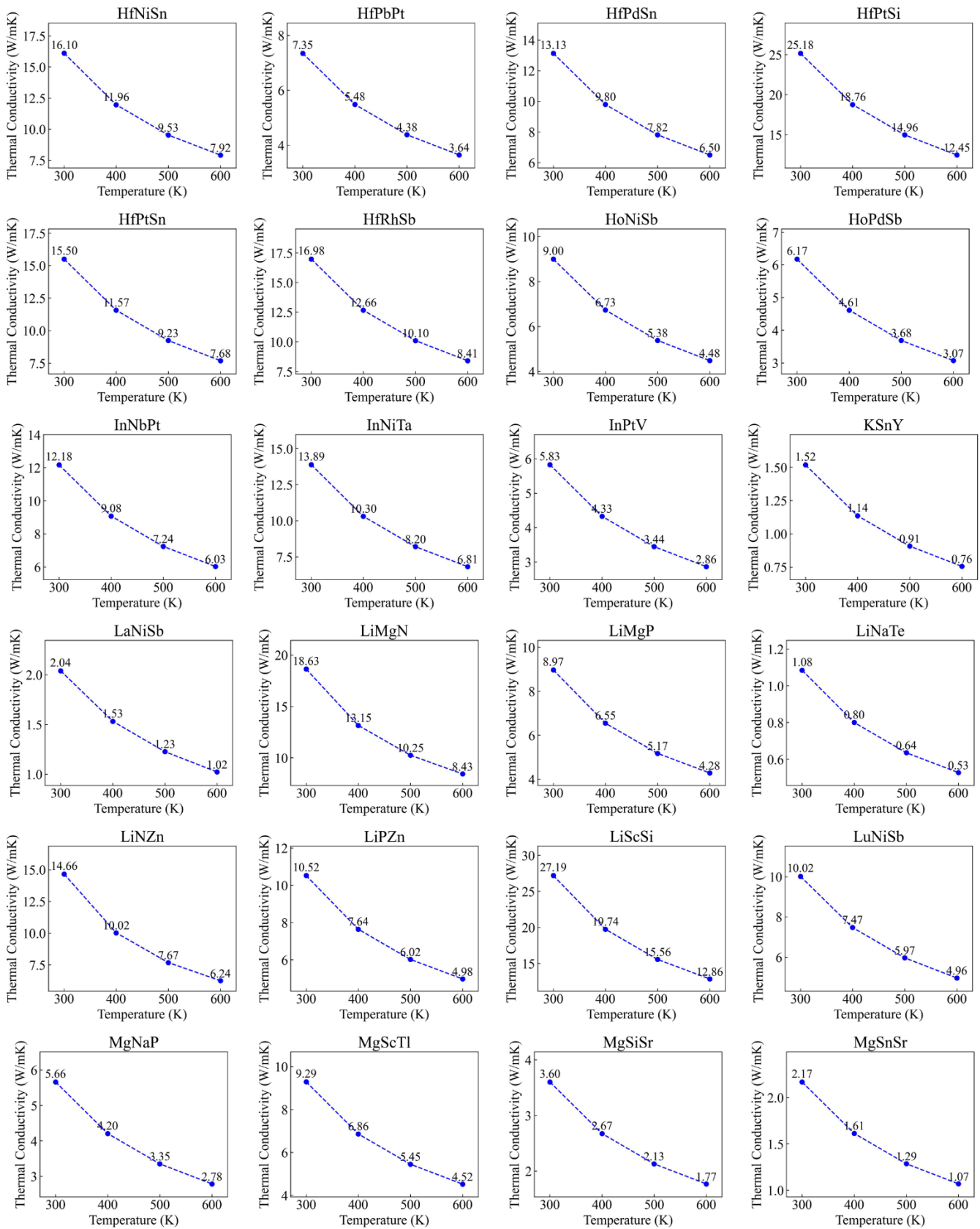

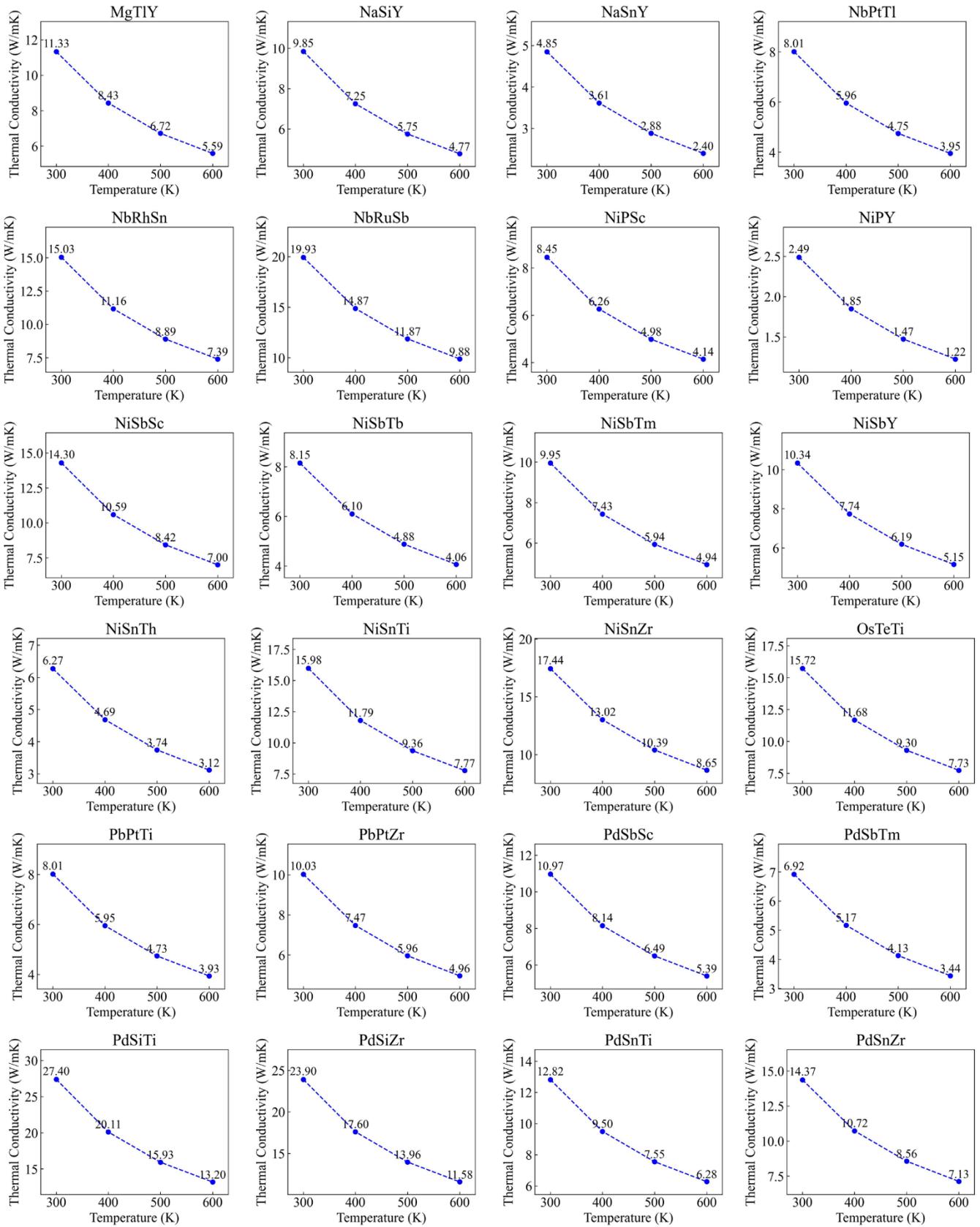

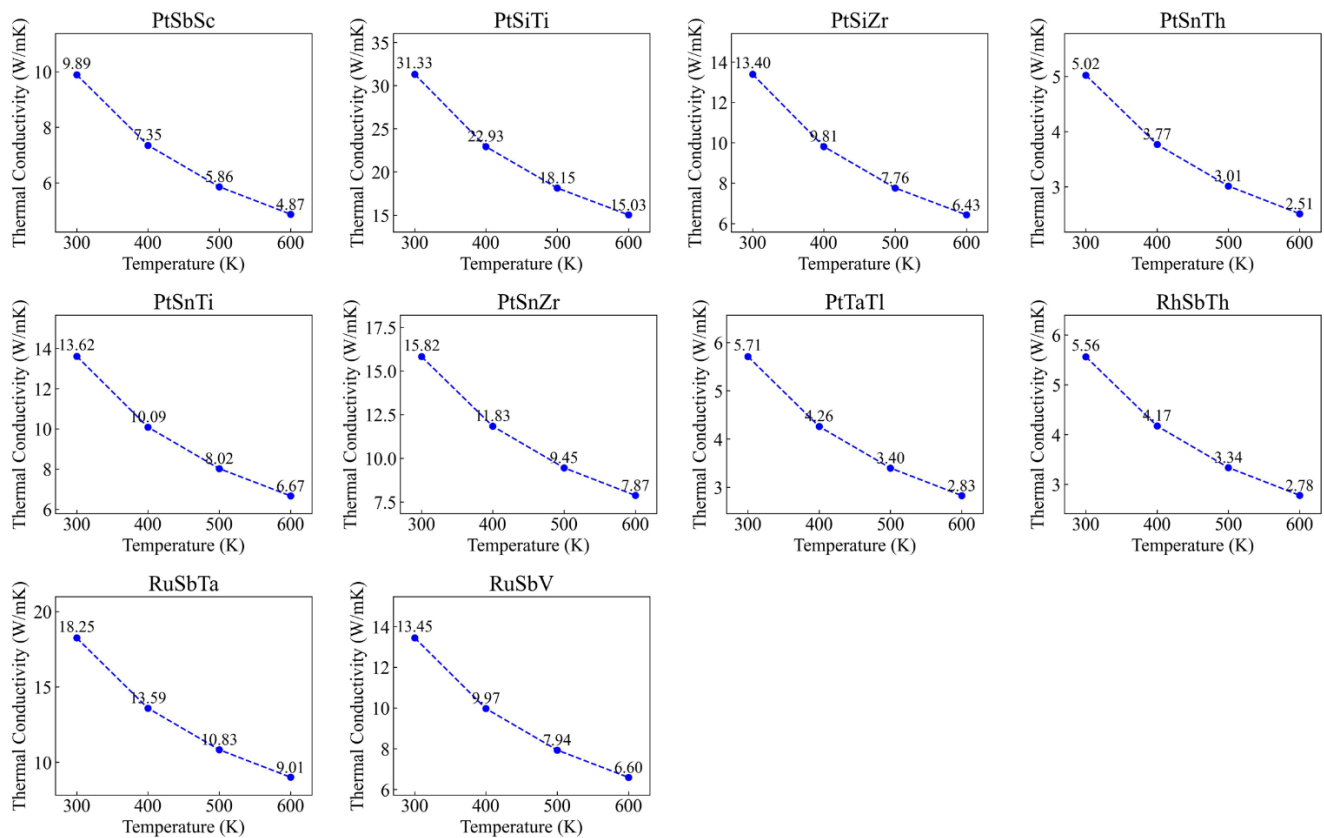

**Fig. S3. Thermal conductivity of 130 half-Heusler compounds.**

## S5. Molecular dynamics simulation of Cu₂Se

All the molecular dynamics simulations in this work were implemented by LAMMPS[4] code.

In the thermal transport simulation of β-Cu₂Se, a 5×5×5 supercell containing 1500 atoms was used. Firstly, a 200 ps NPT simulation and a followed 200 ps NVE simulation with time step of 1fs were performed to equilibrate system. Then, the heat flux was collected every 10 fs during the next 2 ns. For each temperature, 20 independent simulations were performed to obtain the sampling average as final results and the average running thermal conductivities were converged in the time interval [50 ps, 100 ps]. Fig. S4 shows the running thermal conductivity, demonstrating our simulations has reached convergency.

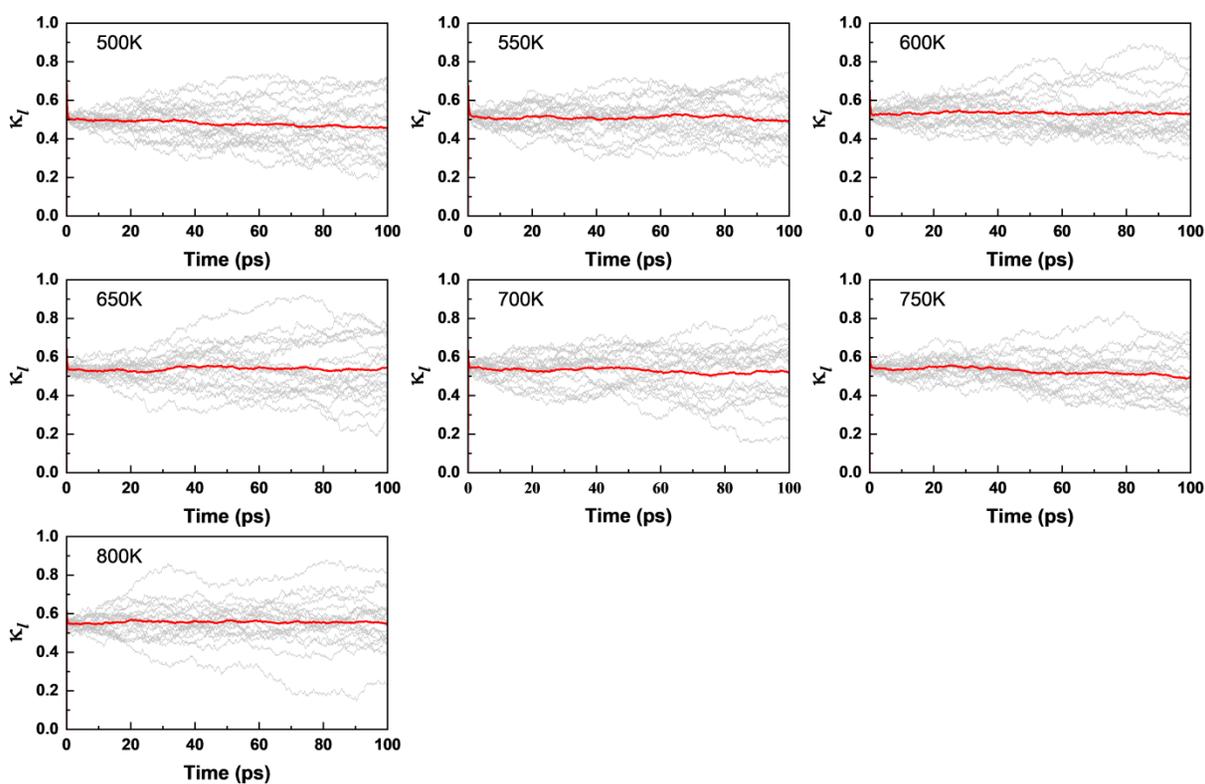

**Fig. S4 Running thermal conductivity of Cu₂Se.** The red line highlights the average result of 20 independent simulations.

In the diffusion simulations, the same box was used. Firstly, a 50 ps NPT simulation was performed to obtain the equilibrium structure at specific temperature. Then a 1ns NVT simulation was performed to obtain the MSD of Cu ions. The MSD data was collected every 0.5 ps. The temporal evolution of Cu MSD is illustrated in Fig. S5.

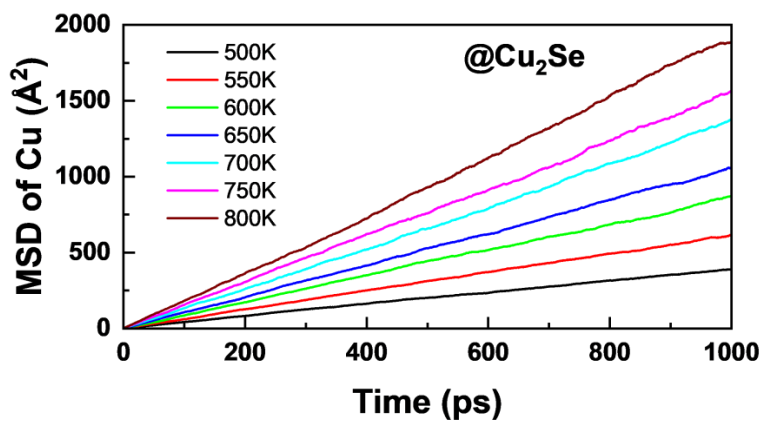

**Fig. S5 Time dependent MSD of Cu at different temperatures in Cu$_2$Se.**

## S6. Molecular dynamics simulation of Sulfide Solid-state Electrolytes

In the diffusion simulation of Li$_{10}$XP$_2$S$_{12}$ (X=Si, Ge, Sn) solid-state electrolytes, a 3×3×3 supercell containing 1350 atoms was used. Firstly, a 50 ps NPT simulation was performed to obtain the equilibrium structure at specific temperature. Then a 1ns NVT simulation was performed to obtain the MSD of Li ions. The MSD data was collected every 0.5 ps. All the molecular dynamics simulations were performed with a time step of 1 fs. The temporal evolution of Li MSD is illustrated in Fig. S6.

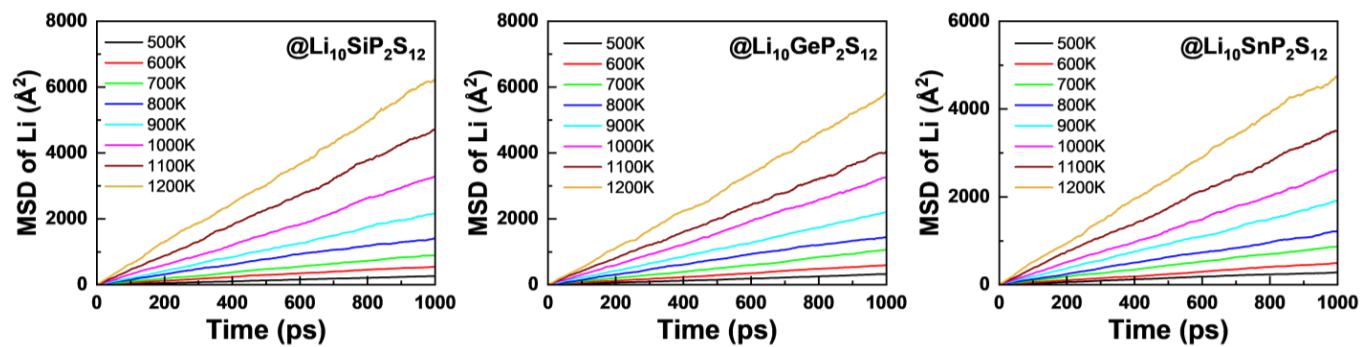

**Fig. S6 Time dependent MSD of Li at different temperatures in Li$_{10}$XP$_2$S$_{12}$ (X=Si, Ge, Sn).**

# Supplementary Reference